\documentclass[superscriptaddress,amsmath,amssymb,notitlepage]{revtex4-1}
\usepackage{graphicx}
\usepackage{dcolumn}
\usepackage{bm}
\usepackage{hhline}

\newcommand{\unit}[1]{\,\mbox{#1}}
\newcommand{\Hz}{\unit{Hz}}
\newcommand{\kHz}{\unit{kHz}}
\newcommand{\MHz}{\unit{MHz}}
\newcommand{\GHz}{\unit{GHz}}

\newcommand{\torr}{\unit{torr}}

\newcommand{\m}{\unit{m}}
\newcommand{\cm}{\unit{cm}}
\newcommand{\mm}{\unit{mm}}
\newcommand{\um}{\unit{$\mu$m}}
\newcommand{\nm}{\unit{nm}}
\newcommand{\K}{\unit{K}}
\newcommand{\mK}{\unit{mK}}

\newcommand{\us}{\unit{$\mu$s}}
\newcommand{\ns}{\unit{ns}}

\newcommand{\degree}{\mbox{$^{\circ}$}}
\newcommand{\degC}{\mbox{\degree{}C}}

\newcommand{\V}{\unit{V}}

\newcommand{\dB}{\unit{dB}}

\newcommand{\W}{\unit{W}}

\newcommand{\pers}{\unit{s$^{-1}$}}

\newcommand{\eV}{\unit{eV}}

\newcommand{\N}{\unit{N}}
\newcommand{\mJ}{\unit{mJ}}

\newcommand{\JHz}{\unit{J/Hz}}
\newcommand{\pscm}{\unit{cm$^{2}$s$^{-1}$}}
\newcommand{\mpers}{\unit{ms$^{-1}$}}

\newcommand{\ff}{\unit{fF}}

\newcommand{\nH}{\unit{nH}}

\newcommand{\kohm}{\unit{k$\mathrm{\Omega}$}}
\newcommand{\C}{\unit{C}}
\newcommand{\pFm}{\unit{pFm$^{-1}$}}

\newcommand{\uG}{\unit{$\mu$G}}

\newcommand{\etal}{{\em et al.}}

\newcommand{\beq}{\begin{equation}}
\newcommand{\eeq}{\end{equation}}

\begin{document}
\DeclareGraphicsExtensions{.pdf,.png,.jpg,.eps,.tiff}

\title{Experimental system design for the integration of trapped-ion and superconducting qubit systems}

\author{D. De Motte}
\affiliation{Department of Physics and Astronomy, University of Sussex, Brighton, BN1 9QH, UK}
\author{A. R. Grounds}
\affiliation{Department of Physics and Astronomy, University of Sussex, Brighton, BN1 9QH, UK}
\author{M. Reh\'{a}k }
\affiliation{Department of Experimental Physics, Comenius University, SK-84248 Bratislava, Slovakia.}
\author{ A. Rodriguez Blanco}
\affiliation{Department of Physics and Astronomy, University of Sussex, Brighton, BN1 9QH, UK}
\author{B. Lekitsch}
\affiliation{Department of Physics and Astronomy, University of Sussex, Brighton, BN1 9QH, UK}
\author{G. S. Giri}
\affiliation{Department of Physics and Astronomy, University of Sussex, Brighton, BN1 9QH, UK}
\author{P. Neilinger}
\affiliation{Department of Experimental Physics, Comenius University, SK-84248 Bratislava, Slovakia.}
\author{G. Oelsner}
\affiliation{Leibniz Institute of Photonic Technology, P.O. Box 100239, D-07702 Jena, Germany.}
\author{E. Il’ichev}
\affiliation{Leibniz Institute of Photonic Technology, P.O. Box 100239, D-07702 Jena, Germany.}
\affiliation{Novosibirsk State Technical University, 20 K. Marx Ave., 630092 Novosibirsk, Russia.}
\author{M. Grajcar }
\affiliation{Department of Experimental Physics, Comenius University, SK-84248 Bratislava, Slovakia.}
\author{W. K. Hensinger}
\affiliation{Department of Physics and Astronomy, University of Sussex, Brighton, BN1 9QH, UK}
\email{W.K.Hensinger@sussex.ac.uk}

\begin{abstract}

We present a design for the experimental integration of ion trapping and superconducting qubit systems as a step towards the realization of a quantum hybrid system. The scheme addresses two key difficulties in realizing such a system; a combined microfabricated ion trap and superconducting qubit architecture, and the experimental infrastructure to facilitate both technologies. Developing upon work by Kielpinski et al. \cite{kielpinski2012}, we describe the design, simulation and fabrication process for a microfabricated ion trap capable of coupling an ion to a superconducting microwave $LC$ circuit with a coupling strength in the tens of kHz. We also describe existing difficulties in combining the experimental infrastructure of an ion trapping setup into a dilution fridge with superconducting qubits and present solutions that can be immediately implemented using current technology. 

\end{abstract}
\maketitle
\section{Introduction}\label{intro}

The advancement of quantum technologies over recent decades has led to the realization of qubits in various physical systems, such as ions \cite{Haeffner2008,Blatt2008}, molecules \cite{Gershenfeld1997}, neutral atoms \cite{Saffman2010}, superconducting qubits \cite{You2011,Devoret2013}, quantum dots \cite{Kloeffel2013}, nitrogen vacancy centres \cite{Jelezko2006,Childress2013} and photons \cite{Kok2007}. Some systems posses natural advantages over others in the form of longer qubit coherence times or faster logic gates. This has led to the promising approach of developing a scalable quantum computer, not only by improving the properties of an individual qubit type, but by exploiting the strengths of different systems to perform different tasks in a combined quantum hybrid system \cite{schoelkopf2008,Kurizki2015}.

Electronic spin systems, such as ions, are well-known for their low interaction with the environment. In particular trapped atoms possess much longer coherence times than superconducting qubits and other solid state systems such as NV centres and quantum dots \cite{Ladd2010,Kurizki2015}. Coherence times are also comparable with photon and nuclear spin systems, such as dopants in Silicon \cite{Kurizki2015}. Ions do however have longer gate times compared with systems such as superconducting circuits and as such it is interesting to explore the use of hybrid systems where one could make use of the fast gate speeds that can be achieved with superconducting qubits. The realisation of such a quantum hybrid system may also have interesting applications for the study of the foundations of quantum physics. 

Superconducting circuits (SC) have shown faster gate times and while they require expensive equipment, such as dilution fridges, they do not require other ancillary equipment such as lasers or magnetic fields, and with microfabrication technologies they lend themselves to scalable architectures. SC can also be grouped together in quantum buses which can process quantum information. SC however have short coherence times and as such they are not ideally suited for the storage of quantum information\cite{Ladd2010,Kurizki2015,schoelkopf2008}.

Two key tasks required for quantum computation are the processing and storage of information. In a quantum hybrid system a dedicated system can be used to perform quantum computation processes, such as quantum logic gates, and a separate system can be used to store of the information. An ideal quantum storage system can be formed from a physical system with coherence times much longer than gate operation times, such as atoms, ions or spins. Likewise, quantum processors can be formed from systems that are capable of performing gate operations at short time scales, such as superconducting qubits.

A major challenge in realizing a hybrid system is the ability to transfer information between the different qubit types. In the case of atoms, ions or spins directly coupling to superconducting circuits via their electromagnetic fields, the coupling is relatively weak due to the systems being far off resonance with each other. 

Ions and superconducting circuits each have properties that we desire to create a functional scalable quantum computer. These systems could be coupled together using a microwave CPW resonator but due to the very low coupling rates ($<1$ Hz) one would need a very large quantity of ions in excess of $N = 10^4-10^6$ which remains a challenging task \cite{Wallquist,Kurizki2015}. By using the electric dipole coupling with Rydberg atoms one could obtain strong coupling ($1-10$ MHz) but such experiments would be tremendously complex due to the experimental difficulty of preparing and manipulating Rydberg ions \cite{schoelkopf2008,Kurizki2015}.

Using the method outlined by Kielpinski \cite{kielpinski2012} one could achieve coupling in the range  $g = 2\pi\times60$ kHz, this is on the order of ion-ion coupling and would therefore allow high-fidelity coherent operations \cite{kielpinski2012,Leibfried2003,Benhelm2008}.

For long range communication it is also possible to integrate photon coupling to the ion for transmission to another dilution fridge with a second ion-SC setup, but this is not discussed in this paper. While the fabrication of the cryogenic BAW described in \cite{kielpinski2012} can be complex, we discuss it's feasibility using existing microfabrication technologies. 

\subsection{Superconducting qubits}

Due to their macroscopic size of 100\um - 1\mm, SC qubits posses large electric dipole moments and for this reason they couple strongly to the microwave field of a CPW or LC resonator. Schoelkopf \etal \cite{schoelkopf2008} have shown the coupling of different quantum systems to microwave photons in transmissions lines. It can be seen that the coupling becomes stronger with the size of the system under consideration and it is higher for systems that posses electric dipole moments compared with magnetic dipole moments.

While superconducting qubits have short coherence times of $\sim$\us, they still have significant potential because of their strong electromagnetic interactions and fast gate operation times of 10's\ns~\cite{Devoret2013}. Superconducting charge and phase qubits couple to the electric field of superconducting resonators, while the flux qubit couples to the resonator's magnetic field \cite{You2011}. For the case of a charge qubit coupled to a single electron field mode, the coupling strength $g$ can be given as \cite{blais2004cavity} 

\begin{equation}
g=\frac{\beta e}{\hbar}\sqrt{\frac{\hbar \omega_r}{cL}},
\end{equation}

\noindent where $L$ is the length of the transmission line resonator, $c$ is the capacitance per unit length of the transmission line resonator, $\omega_r$ is the resonant frequency of the resonator and $\beta \equiv C_g/C_\Sigma$, where $C_g$ is the capacitance of the superconducting island and $C_\Sigma=C_J+C_g$, where $C_J$ is the capacitance of the Josephson junction. For an experimentally realistic system, $\beta \sim 0.1$, $\omega_r \sim 10$\GHz, $L \sim 1$\cm~and $c \sim 50$\pFm, resulting in a coupling strength of $g/2\pi = 50$\MHz \cite{blais2004cavity}.

\subsection{Trapped ions}

Trapped and laser cooled ions have been extensively used to demonstrate many aspects required for quantum computing \cite{Haeffner2008,Blatt2008} and are currently a leading technology with which a large-scale quantum computer can be constructed. Ions trapped in an ultra-high vacuum (UHV) weakly interact with their environment and are easily manipulated using electromagnetic fields. While two-qubit gate operations can take 10-100\us~to perform, trapped ions make ideal systems for storing quantum information with demonstrated coherence times as long as tens of seconds \cite{Ladd2010,Kurizki2015}. 

\paragraph{Magnetic dipole coupling -}Developing upon the work by Verd{\'u} \etal~\cite{verdu2009strong}, who demonstrates strong magnetic coupling of an ultracold atomic ensemble to a superconducting CPW, we can also calculate the expected coupling between a CPW and the hyperfine ground state of an ion. The coupling between a single $^{171}$Yb$^{+}$ ion and a single microwave photon field of a CPW is given as \cite{verdu2009strong}

\begin{equation}\label{equjose}
g=\frac{1}{\sqrt{2}}
\left<1\left|\mu_B
\left(g_s\vec{S}-\frac{\mu_N}{\mu_B}g_I\vec{I}\right)  
\right|0\right>\vec{B}_{trans}\left(\vec{r}\right),
\end{equation}

where the term in the bra-ket represents the M1 transition matrix elements between the two hyperfine levels in the ground state of a $^{171}$Yb$^{+}$ ion, $\mu_B$ is the Bohr magneton, $\mu_N$ is the nuclear magneton, $g_s\sim 2$ is the electron $g$-factor, $g_I$ is the nuclear $g$-factor (where $g_I\ll g_s$) and $\vec{B}_{trans}\left(\vec{r}\right)$ is the transverse magnetic field at an ion height $r$.

As the magnetic field strength decreases with ion height as $B\propto r^{-2}$, one potential solution to increase the coupling is to ensure the lowest experimentally possible ion height. However, a major limiting factor for surface electrode ion traps is the minimum ion height ($r$) required above the CPW. This constraint is due to optical access for ionization and Doppler cooling lasers, as well as ion heating, which scales approximately as $r^{-3.5\pm 0.1}$ \cite{deslauriers2006scaling}. Following these constraints we identify a reasonable ion height as 25\um.

The transverse magnetic field of a cavity can be calculated by performing a numerical simulation using a boundary element method. The single photon cavity-ion coupling strength can be simulated such that the total energy within a defined volume around the cavity is equal to the energy of a single photon at 12.6\GHz. For an ion height of $\sim$25\um, an rf electrode separation of $\sim$20\um~is required. Within the centre of this separation, a superconducting microwave cavity with a width of 10\um, a gap of 5\um~on either side and a height of 5\um~was simulated. The field strength at the ion's position was determined to be $\sim$1\uG. Using Eq. \ref{equjose}, this results in a coupling strength between the ion and CPW of approximately 1\Hz. This is significantly lower than the cavity decay rate of $\kappa=\omega_{CPW}/Q \approx 126$\kHz~for a CPW with a quality factor of $\sim$10$^5$ and with a resonant frequency of 12.6\GHz~corresponding to the hyperfine ground state of $^{171}$Yb$^+$. 

A useful modification to further enhance the coupling is to consider an ensemble of ions, where the coupling constant now scales as $g_{ensemble} =\sqrt{N}g$ for $N$ number of two-level systems \cite{dicke1954coherence}. However, even for large ensembles of $10^6$ ions, the effective coupling strength is still limited to $\sim$1\kHz. Despite this improvement, the coupling strength is still significantly lower than the cavity decay rate. While the cavity decay rate can be decreased by increasing the quality factor of the CPW, currently technology is limited to Q values of $5\times 10^6$ \cite{bruno_reducing_2015}. For coupling via the magnetic dipole in this scheme the decay rate of the resonator mode is dominant, which prevents the efficient and coherent information exchange between the systems.


\paragraph{Electric dipole coupling -}One can potentially increase the coupling rate between an ion and a superconducting circuit with the use of Rydberg ions, by considering coupling via the electric dipole moment. Rydberg ions are ions whose outermost electrons have been excited into a high energy state with a large principle quantum number $n$. The highly excited Rydberg state results in a large electric dipole moment, which could be coupled to microwave resonators on the order of MHz as discussed for atoms \cite{sorensen2004capacitive}. 

Despite the possibility of generating stronger coupling, Rydberg ions also suffer from a number of drawbacks. One primary challenge is the complex laser schemes required to reach the Rydberg states in ions. Either a single vacuum ultraviolet (VUV) laser or multiple UV laser sources are required. VUV laser light is strongly absorbed by air and thus requires the setup to be in vacuum. A recent demonstration of a single trapped $^{40}\mathrm{Ca}^+$ ion utilized a laser at 122\nm~to reach a Rydberg state with $n=64$. Additionally, the energy of a photon at 122\nm, given by $E=\hbar \omega \sim$10.2\eV, is much greater than the work function of commonly used surface materials for trap electrodes, such as gold (5.1\eV). This leads to the added difficulty of electrodes charging and perturbing the trapping potential. Small electric fields can easily ionise Rydberg ions making their use challenging within a conventional ion trap.

\paragraph{Motional dipole coupling -}An alternate indirect coupling method has also been proposed by Kielpinkski \etal~\cite{kielpinski2012}, where the ions electric dipole, induced by ion motion, couples to the electric field of a superconducting $LC$ circuit. This method addresses the challenge of coupling two widely off-resonant systems by modulating the superconducting $LC$ circuit oscillating in the GHz regime, $\omega_{LC}$, with a signal at the ions motional frequency, $\omega_{i} \sim 1-10$\MHz. The ion is confined between two capacitor plates linked to the $LC$ circuit. The coupling between the ion and the circuit then becomes resonant when $\omega_{i} \approx \omega_{LC} \pm \nu$, where $\nu$ is the modulation frequency of the capacitance. Spin-motion protocols based on laser interaction \cite{leibfried2003quantum}, microwave \cite{ospelkaus2011} or radio-frequency fields \cite{lake2015generation} allow for the generation of spin-motion entangled states, allowing the information in the motional state to be swapped to the long lived spin state.

The classical dipole interaction energy of the ion placed within the axial electric field of the capacitor, $E_z$, can be given as \cite{kielpinski2012}

\begin{equation}\label{equ}
U=ezE_z=\frac{e\zeta}{r}zV=\frac{e \zeta}{rC}zQ,
\end{equation}

\noindent where $e$ is the charge of an electron, $z$ is the harmonic oscillator length of the trapped ion, $\zeta$ is the dimensionless geometry factor determined by the structure of the electrodes, $r$ is the ion height, $V$ is the voltage between the capacitor plates, $C=C_0\left[1+\eta \,\mathrm{sin}\left(\nu t \right) \right]$ is the total capacitance of the circuit, $C_0$ is the total static capacitance and $Q$ is the total charge of the circuit. Assuming a modulating capacitance to modulate the circuit frequency, Eq. (\ref{equ}) can be given as \cite{kielpinski2012}

\begin{equation}\label{equ2}
U\left(Q,z,t\right)=\frac{e \zeta}{rC_0}\left[1-\eta \, \mathrm{sin}\left(\nu t\right) \right] zQ,
\end{equation}

\noindent where the modulation depth of the capacitor is kept as $\eta\ll1$. By quantizing the $LC$ and ion motion, keeping the modulated capacitance motion classical and assuming the rotating wave approximation, the Hamiltonian of the $LC$-ion system is given as \cite{kielpinski2012}

\begin{equation}
H_{int}=\frac{2i\hbar g_0\eta}{3}e^{-i\Delta t}ab^\dagger + \mathrm{H.c.},
\end{equation}

\noindent where $\Delta = \nu -\left(\omega_{LC}-\omega_i\right)$, $a$ is the annihilation operator of the microwave photon mode, $b^\dagger$ is the creation operator of the ion motional mode and $g_0$ is the coupling constant given as \cite{kielpinski2012}

\begin{equation}\label{equg}
g_0 = \frac{e\zeta z_0 \Delta q_0}{rC_0 \hbar},
\end{equation}

\noindent where $\Delta q_0$ is the zero-point charge fluctuation on the resonator.
Kielpinski \cite{kielpinski2012} describes the case for a $^9$Be$^+$ ion with a secular frequency of $\omega_i=2\pi \times 1$\MHz~in an ion trap with geometry factor of $\zeta=0.25$. The harmonic oscillator length of the ground-state mode can therefore be given as 

\begin{equation}
z_0=\sqrt{\hbar /2m\omega_i}=24\nm,
\end{equation}

\noindent where $m$ is the mass of the ion. A static capacitance of $C_0=46$\ff~and an inductance of $L_0=400$\nH~are chosen to produce a resonant frequency of 1\GHz, with a characteristic impedance of $Z=2.7$\kohm. This results in a zero-point charge fluctuation of \cite{kielpinski2012,devoret1995quantum}

\begin{equation}
\Delta q_0=\sqrt{\hbar /\left(2Z\right)}=1.4\times 10^{-19}\C.
\end{equation}

\noindent This satisfies the Heisenberg uncertainty relationship $\Delta\phi_{0}\Delta q_{0}\geq\hbar/2$

\noindent where 

\begin{equation}
\Delta\phi_{0}=\sqrt{\frac{\hbar Z}{2}} =3.8  \times 10^{-16} \textrm{Wb}.
\end{equation}

Using Eq. (\ref{equg}), a coupling constant of $g_0/2\pi = 200$\kHz~is achieved using the parameters given above. Assuming a capacitance modulation depth of $\eta=0.3$ (whose motivation is described in section \ref{moddepthsec}), this gives an effective coupling strength of $g/2\pi= 60$\kHz, where $g=g_0\eta$. This coupling is much larger than the decoherence rate of 10$^3$\pers~\cite{kielpinski2012}.

While this method may be capable of producing a strong coupling between the ion and the resonator, it remains an experimentally challenging task to incorporate all necessary features onto a single microfabricated ion trap. In section \ref{bawsec} we discuss how this can be accomplished and present the design, simulation and microfabrication methods for an ion trap incorporating all features to realize this scheme.  

The development of an experimental infrastructure capable of operating both systems also provides significant challenges. While some requirements for both systems are similar, such as low noise and the use of vacuum systems, the operation of superconducting qubits requires the number of quasiparticles in the superconductor to be negligible. Thus, the temperature of the system has to be kept under 50\mK~for aluminium technology, where the Josephson junction is formed from an aluminium oxide layer in between two layers of aluminium. External magnetic fields must also be kept lower than the critical magnetic field of the superconductor.

The only technology that can currently cool macroscopic objects to temperatures lower than 100\mK~is a dilution refrigerator. The requirements for ion trapping, such as laser beams, electrical connections, atomic source ovens and ion imaging, must therefore be made compatible with the dilution refrigerator environment. The cooling power of commercially available dilution refrigerators (on its mixing chamber) is no more than $\sim$10 mW at 100\mK. The power dissipation of all devices mounted on mixing chamber plate must therefore be lower than this value while retaining sufficient radiation shielding and thermal insulation. In particular, the heat generated by the application of the trapping RF should be minimized. This can be achieved with the use of high Q resonators and low loss dielectric materials. In section \ref{dilsec} we describe the experimental challenges involved in incorporating and operating an ion trapping experiment in a dilution fridge, and propose solutions that can be implemented using current technology. 

\section{Ion trap and $LC$ circuit}\label{bawsec}

\begin{figure}[t!]
  \begin{minipage}[c]{0.5\textwidth}
    \includegraphics[width=\textwidth]{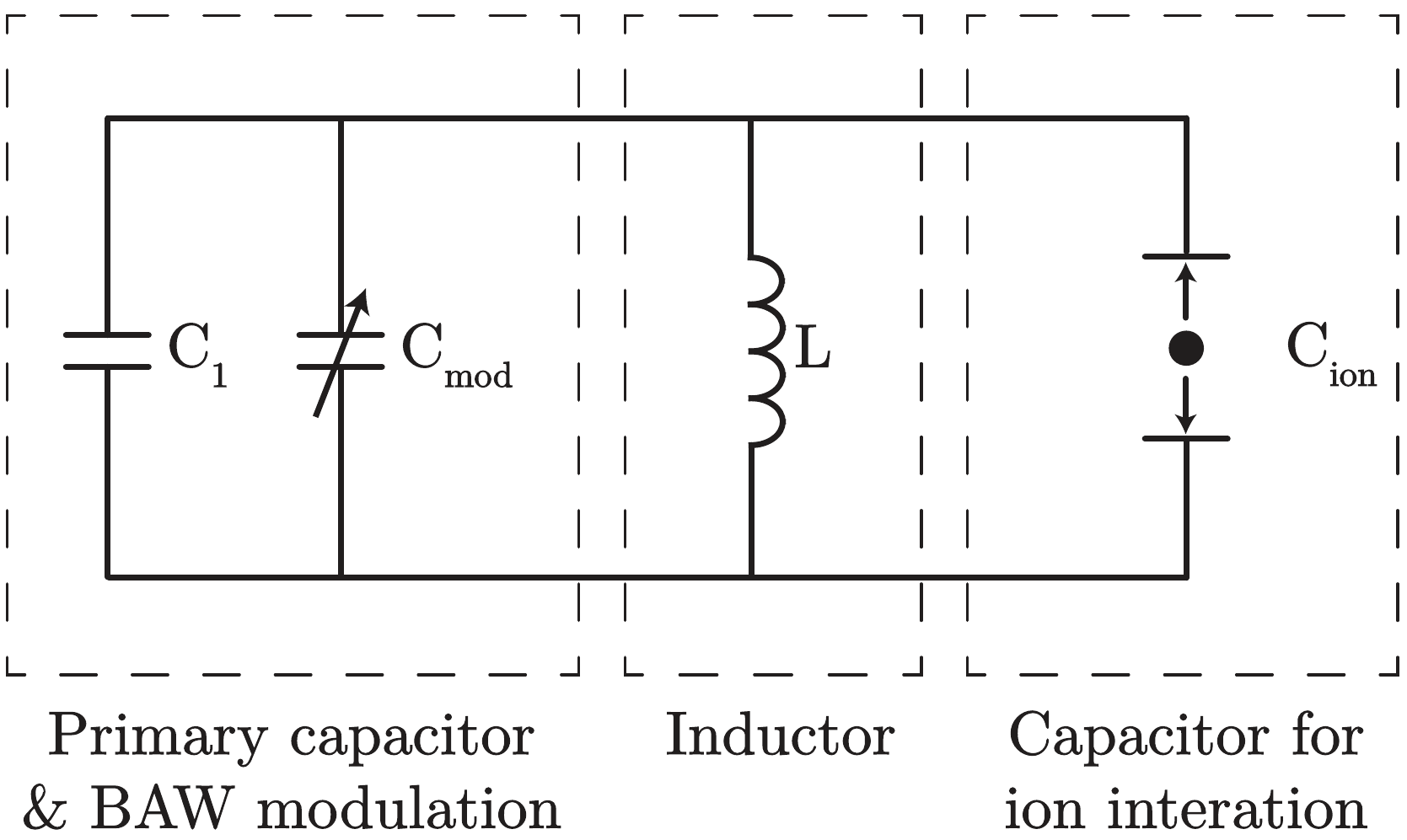}
  \end{minipage}\hfill
  \begin{minipage}[c]{0.4\textwidth}
    \caption{Circuit diagram for a superconducting $LC$ circuit with a resonant frequency of $\sim$1\GHz. A modulated capacitance provides a frequency modulation of $\sim$1\MHz~using a BAW resonator. A secondary capacitor is used to directly couple the motion of the ion with the electric field of the capacitor.}
     \label{circuitdiagram}
  \end{minipage}
\end{figure}

To create the hybrid device proposed by Kielpinski \etal~\cite{kielpinski2012}, the complete structure can be first divided into its individual components, where each component performs a specific function. The first feature to be considered is the ion trapping architecture. For surface electrode ion traps, this is formed from two parallel electrodes which generate a pseudopotential from an applied rf voltage. Additional electrodes with dc voltages are also used to provide axial confinement, shuttling capabilities and micromotion compensation \cite{nizamani2012optimum}.

The superconducting $LC$ circuit is formed from four individual components; an inductor, a primary capacitor, a secondary capacitor for ion interaction, and a bulk acoustic wave resonator (BAW). Fig. \ref{circuitdiagram} shows the circuit diagram used to model the system, where the BAW is used to modulate the capacitance of the primary capacitor at the ions secular frequency. The BAW itself is formed from a piezoelectric material sandwiched between two conducting layers. Applying a voltage to the conducting plates, generates an acoustic wave which travels through the bulk of the material. Depending on the geometry and material of the piezoelectric layer, the resonant frequency of the device can be determined.

BAWs originally gained popularity as filters \cite{aigner2005mems} and are now also commonly used as sensors \cite{johnston2010fbar} and oscillators \cite{petit2008thermally}. They can be designed and microfabricated into two distinct topologies, as film bulk acoustic resonators (FBAR) or solidly mounted resonators (SMR). FBAR topologies are fabricated such that the BAW is surrounded by an air gap. This is accomplished either by fabricating the BAW upon a sacrificial layer or by backside etching directly through the wafer (see Fig. \ref{baw}). SMRs substitute the air gap with alternating layers of different acoustic velocities. Each layer posses a thickness equal to a quarter wavelength with the stack of layers forming a Bragg acoustic reflector.


\subsection{Design and simulation}\label{designandsim}

\subsubsection{Materials}
The first requirement of the simulation process is to determine the materials required for each part of the device. For the piezoelectric material there are several options that can be used in a microfabrication process. Aluminium Nitride (AlN), Gallium Nitride (GaN), Zinc Oxide (ZnO) and Lead zirconate titanate (PZT) have all been demonstrated for BAW resonators. While AlN and PZT are the most commonly used materials, offering well established fabrication procedures, PZT was ultimately chosen for this proposal due to its lower acoustic velocity (4600\mpers) compared with AlN (11,000\mpers). The lower acoustic velocity results in shorter travelling wavelengths through the material for a given frequency of operation, allowing for smaller feature sizes.

\begin{figure}[t!]
\includegraphics[width=1\textwidth]{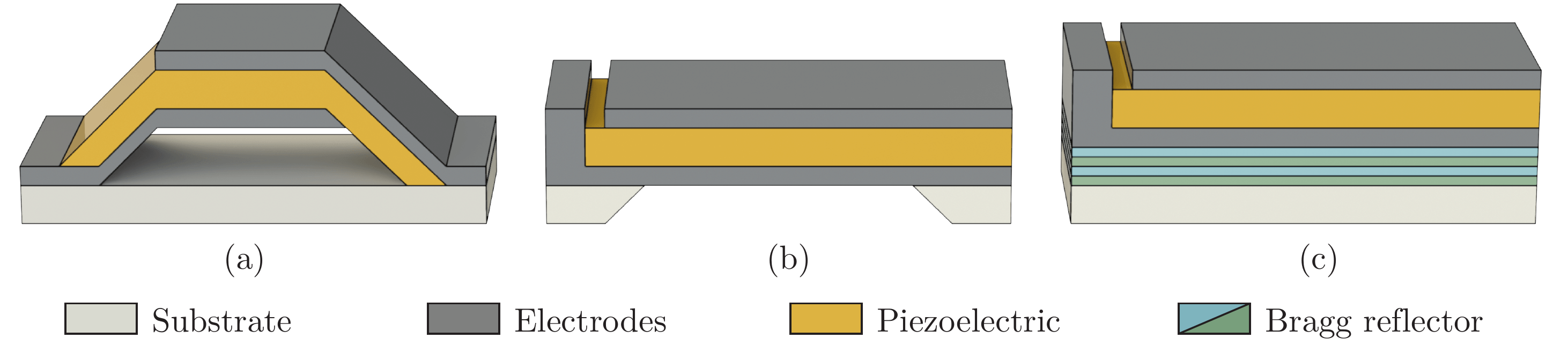}
\caption{(a) An arched film bulk acoustic resonator (FBAR) structure fabricated by depositing the BAW device on top of a sacrificial layer which is later removed to leave an air gap. (b) A flat FBAR structure fabricated by backside etching through the wafer. (c) A solidly mounted resonator (SMR) structure with an alternating layer Bragg reflector stack.}
\label{baw}
\end{figure}

For the ion trapping electrodes and the $LC$ circuit, a superconducting material must be chosen. 
Typically for superconducting qubits the resonator material is formed from Niobium (Nb), a type-II superconductor which has a critical temperature of 9.3\K. Nb has also been demonstrated as an electrode material for surface electrode ion traps \cite{wang2010superconducting} and therefore presents itself as an ideal candidate. For insulating layers used in between conducting layers, silicon dioxide (SiO$_2$) and silicon nitride (Si$_3$N$_4$) were chosen due to their demonstration in both BAW \cite{omori2001pzt, wang2013bending} and ion trap structures \cite{sterling2013increased}.

Finally, sapphire (Al$_2$O$_3$) was chosen as the wafer material as it posses a low microwave loss tangent ($10^{-4}$). Sapphire also possesses a high dielectric constant of 11.5, which results in shorter wavelengths for a given frequency within the $LC$ circuit, allowing for the further reduction in size of the circuit.

\subsubsection{The ion trap}
The design and microfabrication of surface electrode ion traps (including those with superconducting electrodes) has been well described and demonstrated  \cite{hughes2011microfabricated, seidelin2006microfabricated, chiaverini2005surface, antohi2009, labaziewicz2008suppression}. An ion height of 25\um~is chosen by Kielpinski \etal~\cite{kielpinski2012} as a minimum based on current achievements for ion trapping on surface electrodes ion traps \cite{seidelin2006microfabricated, ospelkaus2011}. To create the trapping potential, two rf electrodes with widths of $b$ and $c$ and a separation $a$ are required. For efficient Doppler cooling of the ion in all three axes, the laser wave vector $k$ must have a component along all three principal axes of the ions motion. This can be achieved by making the rf electrodes with differing widths such that $b \neq c$. For $c=b/2$, the optimum ratio of rf electrode separation is given by $\zeta=b/a=4.90$ \cite{nizamani2012optimum}. Using the dimensions of $a$, $b$ and $c$, the ion height, $h$, can be expressed as \cite{nizamani2012optimum}

\begin{equation}\label{height}
h=\frac{\sqrt{abc\left(a+b+c\right)}}{b+c}.
\end{equation}

For an ion height of 25\um, $a$, $b$ and $c$ can be chosen as 18\um, 90\um~and 45\um~respectively, satisfying Eq. (\ref{height}) where $\zeta=4.90$ and $c=b/2$. Static dc potential electrodes can also be placed in between and on the outside of the rf electrodes. Optimum dc electrode widths can be determined as a function of rf electrode separation $a$ \cite{nizamani2012optimum}. 

For an outer segmented electrode chip the width of the electrodes is chosen to be $w_{outer}\approx 3.66a=66 \mu m$ \cite{nizamani2012optimum} in order to maximize the quartic term $\beta$ of the electrical potential Taylor expansion, near the centre of the trap in the axial z-direction $V(z)\approx 2e\gamma z^2+2e\beta z^4$ (with $\gamma<0$ and $\beta>0$ to create a potential wedge for ion separation processes). Ion separation is a useful feature for proposed large scale ion trap quantum computers, which possess trapping, detection and interaction zones \cite{lekitsch2015blueprint}.



\subsubsection{The bulk acoustic wave resonator}

From the three BAW resonator designs presented in Fig.~\ref{baw}, an arched FBAR structure was chosen for this proposal as it provides two key benefits. Firstly, the removal of a single $2-5$\um~layer using a wet etching process offers the simplest fabrication process. This helps to minimize fabrication errors when compared with backside etching through $\sim$500\um~of wafer material for a flat FBAR, or the repeated deposition of multiple Bragg reflector layers for an SMR. Secondly, FBAR structures require materials only for electrode and piezoelectric layers, unlike SMRs which require two additional materials for the reflector stack.

\begin{figure}[t!]
\includegraphics[width=1\textwidth]{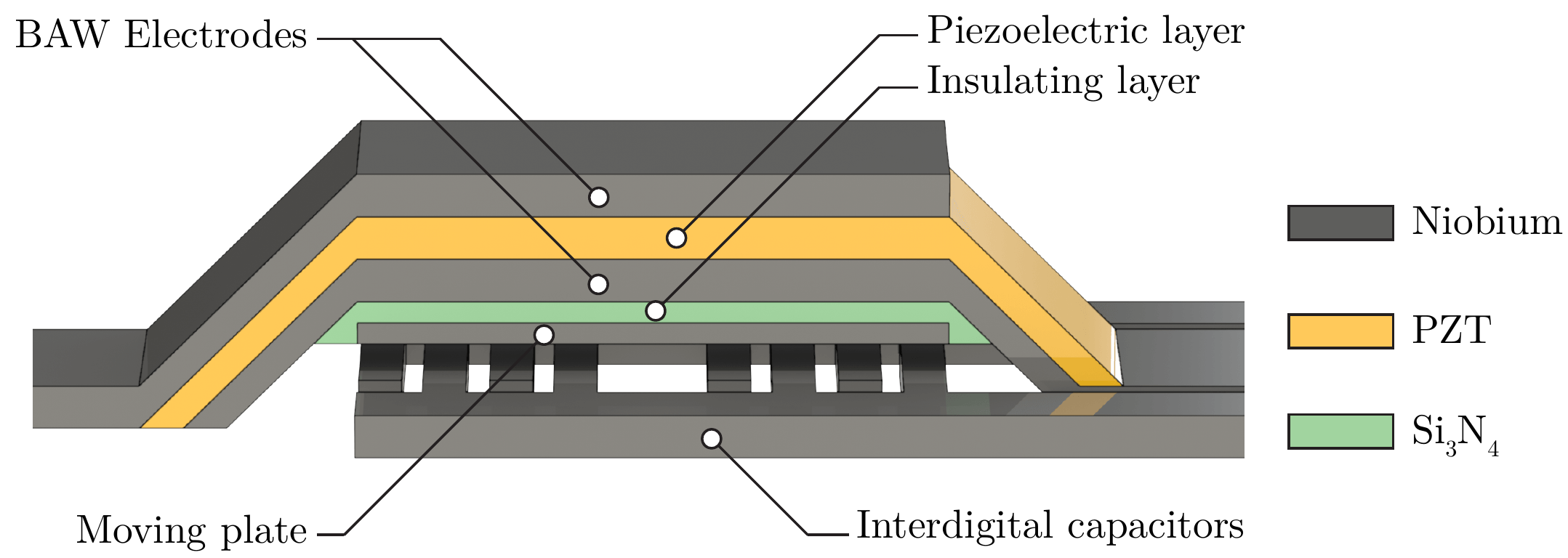}
\caption{An arched film bulk acoustic resonator (FBAR) with a PZT piezoelectric layer and Nb electrode layers. An additional Si$_{3}$N$_{4}$ and Nb layer are placed on the underside of the resonator, with the Nb layer serving a moving plate that modulates the capacitance of the primary capacitor.}
\label{bawclose}
\end{figure}

For many BAW applications, thin films of PZT in the order of a few hundred nanometres can be sufficient. However, for actuation purposes it is preferable to have thicker layers to obtain structural stability during flexing. For this reason, a 3\um~thick PZT layer was included in this proposal as this is the maximum thickness achievable from current sputter deposition processes \cite{wilke2013sputter}. A 1\um~thick Nb layer was added to the top and bottom of the PZT to form the electrode structures (see Fig. \ref{bawclose}). Due to the thickness of the PZT and Nb layers, stresses due to resonator flexing was found to be negligible during simulation. The amplitude of the PZT oscillation also falls below critical limits for single-crystal microresonators \cite{kaajakari2004}, with amplitudes up to 5\um~being demonstrated for 1-3\um~thick and 700\um~long PZT cantilevers \cite{yeatman1995}.

An additional Si$_{3}$N$_{4}$ and Nb layer was placed on the underside of the resonator to serve as an electrically insulating and conducting layer respectively. The conductive layer serves as an electrically conducting plate near the primary capacitor, which moves with the BAW resonator. In section \ref{moddepthsec}, the primary capacitor in the form of an interdigital capacitor is chosen. The capacitance forms mainly from the surface area along the edges of each digit of the capacitor. This is due to the electric field being confined primarily in the regions between each digit. By introducing an electrically conductive plate above the fingers of the interdigital capacitor, we can change its capacitance. This is achieved by confining the electric field above the top surface of the interdigital capacitor and increasing its relative contribution to the total capacitance. By attaching the conductive plate to the BAW, we can effectively modulate the capacitance of the primary capacitor.

The introduction of a modulating plate above a capacitor with a fixed capacitance also has a number of benefits, when compared to solely modulating the position of one side of the capacitor. In a two-plate capacitor, where the position of a single plate is modulated and the other plate is fixed, the driving signal is coupled strongly to the total capacitance in the circuit. Any noise introduced by the driving signal would also be coupled strongly to the circuit. 

We assume that the primary source of noise in the system will be electrical from outside the system as the temperature is in the mK regime, so thermal noise sources are reduced. The other significant source of noise originates from vibrations due to the operation of the dilution fridge and other apparatus. These dominant forms of noise would affect the modulating capacitor more in a system with low drive signal to modulation depth.

An effective three-plate capacitor design is chosen as the two fixed plates form the bulk of the total capacitance and the third plate forms the modulation. Therefore, any noise on the driving signal is decoupled from the total capacitance and we obtain a larger signal to noise ratio. The insulating layer must be used as the moving plate attached to the BAW forms part of the $LC$ circuit and should be decoupled from any potential sources of noise, such as the driving signal on the BAW.

During the design process it is important to separate the frequency of the available modes in the BAW such that the driving energy only couples to the single mode we require. It should be ensured that the harmonics of the other principal modes such as longitudinal, thickness shear and face shear are not also driven. This can be accomplished by keeping the dimensions of thickness, width and length non-divisible with each other. This ensures that the resonator only resonates at one mode within the driving frequency range with a typical Q for the chosen BAW material and dimensions. The Q of the BAW is high enough that only a single mode is driven.


\begin{figure}[t!]
\includegraphics[width=1\textwidth]{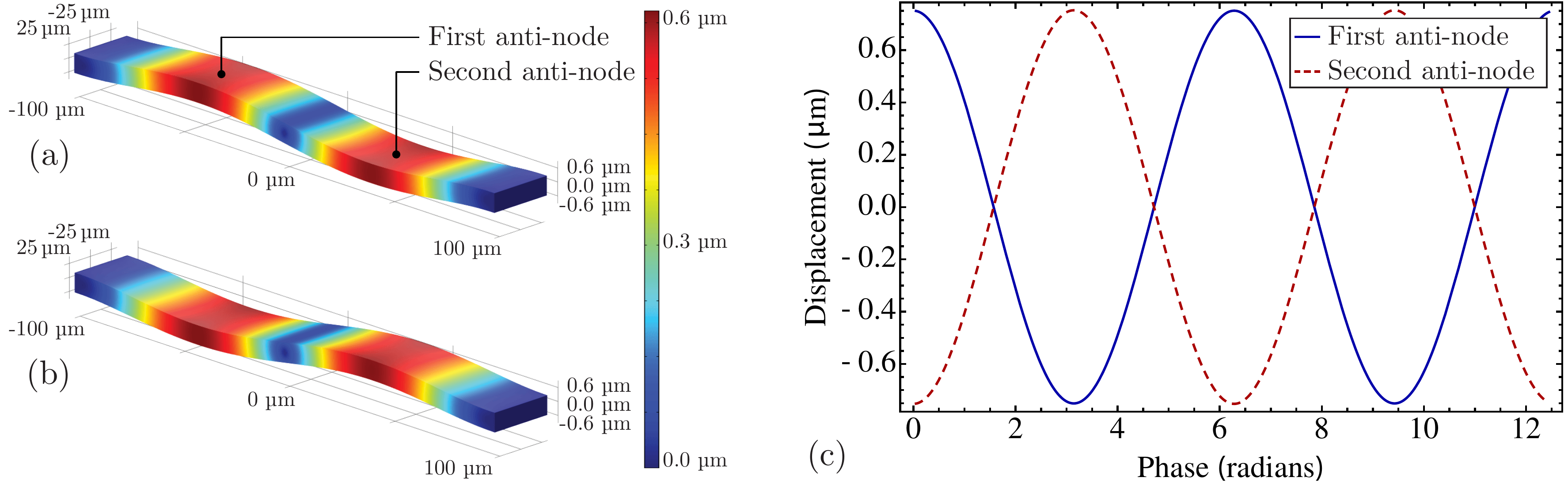}
\caption{A $200\times50\times3$\um~BAW device resonating at a 1\MHz~second harmonic flexure mode, where (a) and (b) show the resonator at a phase of 0 and $\pi$ respectively during its oscillation. The flexure on (a) and (b) is accentuated by a factor of 5. (c) Displacement at the 1$^{\mathrm{st}}$ (solid) and 2$^{\mathrm{nd}}$ (dashed) antinodes of the BAW resonator.}
\label{proxy28}
\end{figure}

While driving the resonator with a sinusoidal signal produces a sinusoidal resonator motion, the variation in capacitance due to the moving plate can contain significant additional harmonics, depending on the modulation depth. The capacitance of the primary capacitor is determined by the separation of the capacitor plates, and varies as $C = C_0 \times \alpha/(\alpha+\beta\mathrm{sin}\left(\omega t\right))\approx C_0(1 + \frac{\beta}{\alpha}\sin{\omega t}+\frac{\beta^2}{\alpha^2}\sin^2{\omega t})$ to second order in the binomial expansion, where $C_0$ is the static capacitance, $\beta$ is the modulation amplitude, $\omega$ is the frequency, and $\alpha$ describes the fixed separation distance when no modulation is present.

This non-sinusoidal capacitance can be partially corrected for by replacing the single primary capacitor with two capacitors connected in parallel and modulated with a $\pi/2$ phase difference. This leads to a capacitance given by

\begin{equation} \label{cap}
\begin{split}
C &= \frac{C_0}{2}(\frac{\alpha}{\alpha + \beta \mathrm{sin}\left(\omega t\right)} + \frac{\alpha}{\alpha + \beta \mathrm{cos}\left(\omega t\right)})\\
 &\approx \frac{C_0}{2}(1 - \frac{\beta}{\alpha}\mathrm{sin}\left(\omega t\right) + +\frac{\beta^2}{\alpha^2}\sin^2{\omega t} + 1 - \frac{\beta}{\alpha}\mathrm{cos}\left(\omega t\right)) + \frac{\beta^2}{\alpha^2}\cos^2{\omega t}\\
&= C_0(1+\frac{\beta^2}{2\alpha^2}) - \frac{\beta}{\alpha}\frac{C_0}{\sqrt{2}}\mathrm{sin}\left(\omega t + \frac{\pi}{4}\right),
\end{split}
\end{equation}

With $\beta << \alpha$ we get

\begin{equation}
C = C_0 - \frac{\beta}{\alpha}\frac{C_0}{\sqrt{2}}\mathrm{sin}\left(\omega t + \frac{\pi}{4}\right).
\end{equation}

The same resonator can be used to modulate both capacitors by driving the resonator on its second harmonic, creating two displacement antinodes. Each capacitor can therefore be coupled separately to the motion at each antinode. Figs. \ref{proxy28}a and~\ref{proxy28}b show the second harmonic flexure mode of the resonator, with the curvature of flexure accentuated by a factor of 5. Fig.~\ref{proxy28}c shows the displacement of the resonator at each antinode with Fig.~\ref{capacitance}b showing the total capacitance for both capacitors modulated in parallel. The final dimensions for the piezoelectric layer are $200\times50\times3$\um, which produces a second harmonic resonant frequency of $\sim$1\MHz.

\subsubsection{The primary capacitor}\label{moddepthsec}

For the circuit design described by Kielpinski \cite{kielpinski2012}, a total capacitance of 46\ff~(coupled with an inductance of 440\nH) is present to produce a resonant frequency of $\sim$1\GHz. From Eq. (\ref{equg}) we see that for lower values of capacitance the coupling also increases. However, the value of 46\ff~presented was limited primarily due to parasitic capacitances originating from the coiled inductor. In section \ref{inductorsec} we describe the use of an alternate inductor design with parasitic capacitances $<$5\ff, which allows us to reduce the total capacitance of the circuit. However, as proof of principle, we maintain a total capacitance of 46\ff~here and present the benefits of a decreased capacitance in section \ref{inductorsec} after all the components of the $LC$ circuit have been described.

\begin{figure}[t!]
\includegraphics[width=1\textwidth]{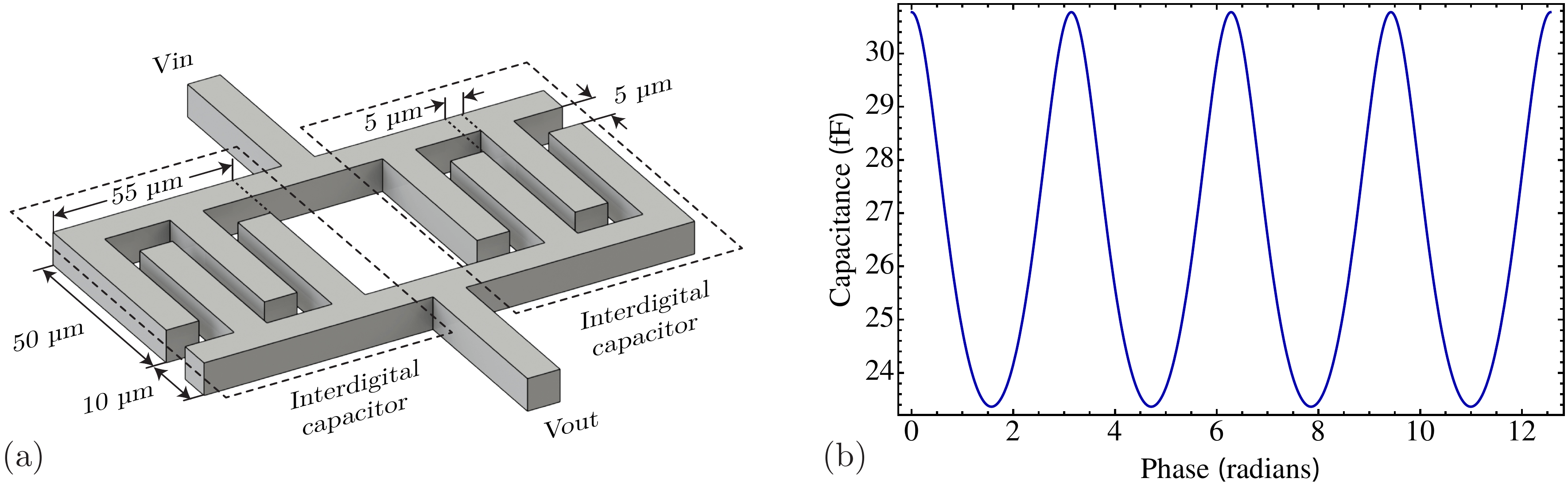}
\caption{(a) Diagram of two four-fingered interdigital capacitors. Each finger has dimensions of $5\times84\times1$\um~with finger spacings of 5\um. (b) Total capacitance of two interdigital capacitors in parallel modulated $\pi$ out of phase with each other.}
\label{capacitance}
\end{figure}

Another parameter that requires definition is the modulation of the capacitor using the BAW, which produces the frequency modulation for the circuit. The frequency modulation takes the form of $\mathrm{cos}\left[\omega_t+f(t)\right]$, where $f(t)$ is the modulation function. If we assume that the modulation is a sine function of a single frequency, we can express the time evolution of the circuit as 

\begin{equation}
y(t)=A_c \, \mathrm{cos} \left[2\pi f_c t+\frac{f_{\Delta}}{f_m} \, \mathrm{cos} \left(2\pi f_m t\right)\right], 
\end{equation}

\noindent where $A_c$ is the carrier amplitude, $f_c$ is the carrier frequency, $f_m$ is the modulation frequency, and $f_{\Delta}$ is the carrier frequency variation under modulation. The fraction $f_{\Delta}/f_m$ is also known as the modulation index or modulation depth. For a carrier modulated by a sinusoidal signal, the resulting frequency spectrum can be derived using Bessel functions of the first kind, as a function of the sideband number and the modulation depth. Ideally, we wish to keep the sideband number to one, confining the energy within carrier and first sideband. Therefore, a modulation depth of $\sim$0.3 is selected as it provides the maximum energy within the first sideband while keeping the majority of power ($\gtrsim 99\%$) within the carrier and first sideband. While the energy within the first sideband can be increased for higher modulation depths, the energy with the carrier decreases such that the combined carrier and first sideband energy is significantly lower. 


To modulate the capacitor effectively, the capacitor should also be placed directly beneath the BAW. Any part of the capacitor outside the BAW only contributes to the total capacitance and not to the modulated capacitance. Therefore, it is beneficial to keep the surface area of the capacitor equal to or smaller than the BAW device. A compact solution for microfabricated devices is an interdigital capacitor that uses edge coupling between tracks to produce a capacitance.

Taking the requirements for total capacitance, modulation depth and spatial area, the digits of the capacitor can be optimized to meet all requirements. The simplest geometry, while maintaining feature sizes $>$5\um~for ease of fabrication, is a four finger geometry. Two four-fingered capacitors in parallel are capable of producing a capacitance of 46\ff, a modulation depth of 0.3 (using a BAW with a separation of 2\um~and oscillating with a 0.6\um~amplitude) and with a complete geometry smaller than the BAW structure. The dimensions of the capacitor are given in Fig. \ref{capacitance}a.


\subsubsection{Capacitor for ion interaction}

When designing the coupling capacitor, we wish to maximise the visible surface area to the ion in order to increase the surface charge affected by the ions motion. As the plate is superconducting, the charge flow will be confined to the surface. By maximising the plate's surface area towards to ion, we also maximise the total number of surface charge carriers and induced surface current density.

\begin{figure}[t!]
\includegraphics[width=1\textwidth]{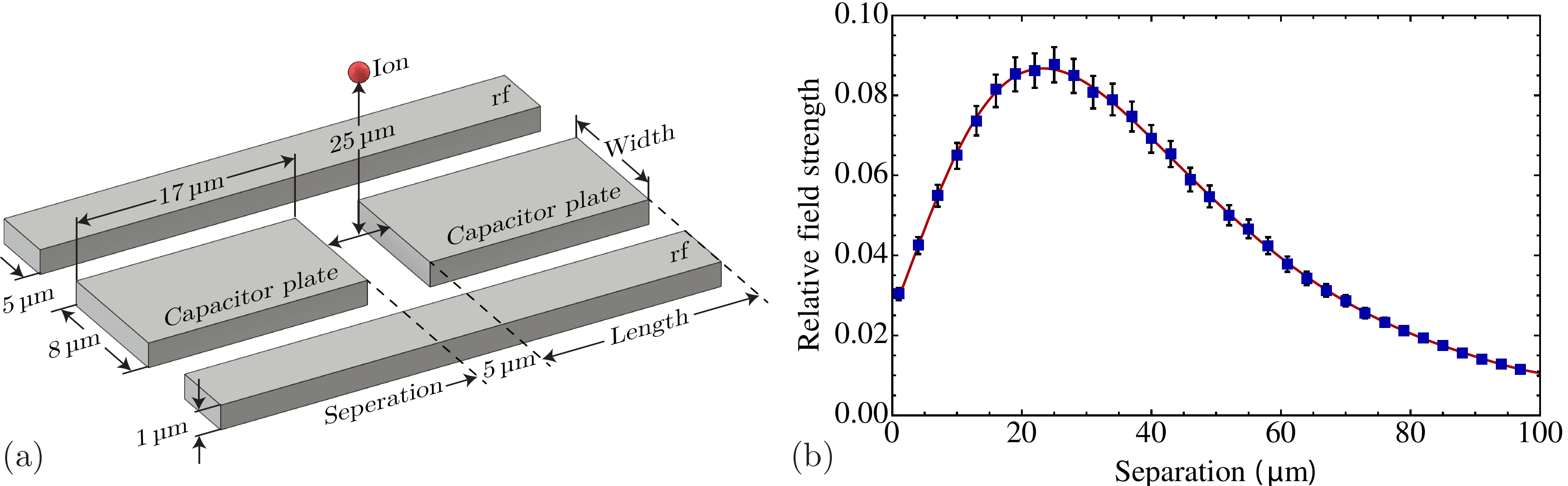}
\caption{(a) Diagram of two capacitor plates used for ion interaction on the surface of the ion trap. Each plate has dimensions of $8\times17\times1$\um~with a 5\um~separation in between both plates. (b) Simulated variation in relative field strength for a fixed charge at an ion height of 25\um~as a function of separation between the centre of two capacitor plates. Error bars represent the uncertainty of the numerical simulation.}
\label{ionpads}
\end{figure}

For the structure of the plates we can begin by assuming a rectangular geometry to maximise the surface area within the rf electrodes (see Fig. \ref{ionpads}a). This gives parameters of length, width and separation, which can be optimized. The coupling capacitance formed to ground must also be considered as this increases with the surface area and therefore limits the maximum size of the plates.

For an ion height of 25\um, an optimum separation of $\sim$24\um~between the centre of each plate was found, where the relative field strength for a given surface charge is maximum at the ions position (see Fig.~\ref{ionpads}b). The field strength at the ion's position also increases linearly with the length of the plate. A length of 17\um~was therefore chosen for each plate, which allows for a 24\um~separation between the centre of each plate while leaving a 5\um~gap in between both plates for ease of fabrication. Finally, the plate widths, which are limited by the rf electrodes separation of 18\um, can be optimized. The maximum width can be determined by selecting an upper limit for the coupling between the plate and the rf electrodes. Here, an upper limit of a 10\% increase in the total circuit capacitance was chosen, resulting in a plate width of 8\um~and a plate-electrode separation of 5\um. Fig.~\ref{ionpads} shows the final geometry and dimensions of the capacitor plates.


\subsubsection{The inductor}\label{inductorsec}
The inductor proposed by Kielpinski \cite{kielpinski2012} is a microfabricated superconducting coil with a 1\mm~diameter and a length of 650\um. While coiled inductors are a suitable solution in macroscopic circuits, they can be difficult to implement into microfabricated systems, requiring a multilayer fabrication process and multiple vias. Inductor coils can also be susceptible to parasitic and self capacitances, introducing noise to the circuit.

\begin{figure}[t!]
    \includegraphics[width=\textwidth]{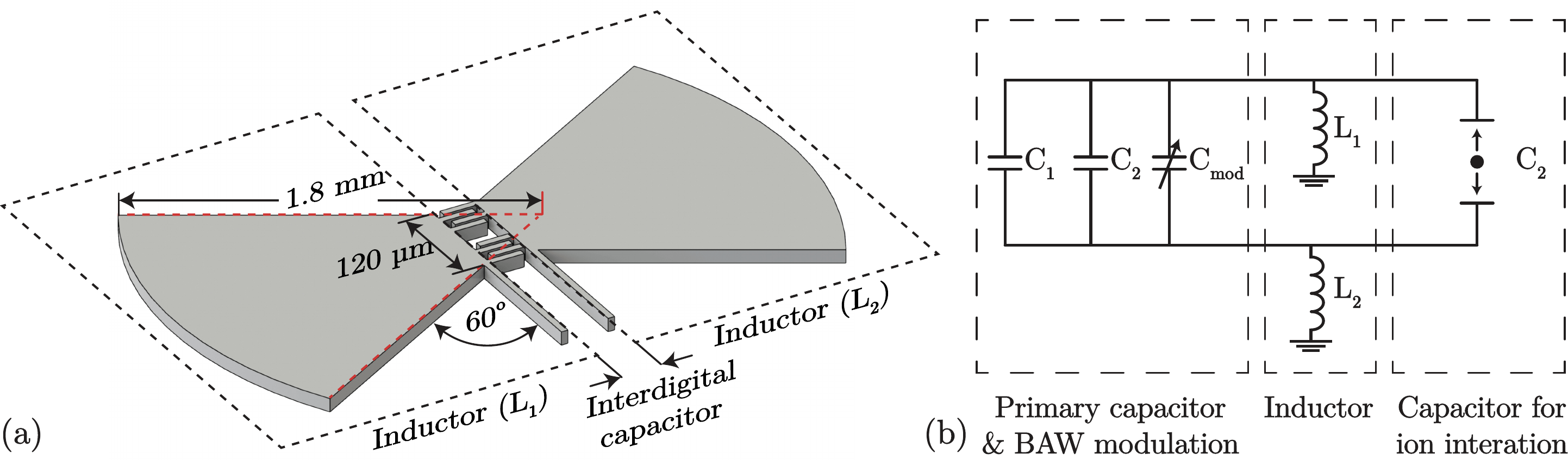}
    \caption{(a) Diagram of two dovetail stub inductors on either side of the primary interdigital capacitor. The inductor posses a 120\um~stub width which fans out at 60$^{\circ}$ to a radius of 1.8\mm. (b) Equivalent circuit diagram for the dovetail inductor geometry and parallel interdigital capacitors.}
     \label{dovetail}
\end{figure}

For microfabricated devices there are several alternative geometries which can produce an inductance, such as spiral, meandering and stub inductors. Microstrip stubs provide the simplest solution and are formed from a straight transmission line connected only at one end, with the free end either open or short circuited. Neglecting losses in the transmission line, the input impedance of the stub is purely reactive. Depending on the electrical length of the stub it can act as either a capacitor or an inductor. As the wave propagates down the stubs length, the stub switches between being inductive and capacitive every $\lambda/4$.

The geometry chosen for simulations was a radial (or ``dovetail'') stub design. As radial stubs fan out, this increases the bandwidth of the stub, reducing end effects such as parasitic electric fields. The larger tail perimeter also results in a lower impedance, which scales down any impedance variation due to microfabrication error. For symmetry about the centre of the primary capacitor, one inductor is used on either side of the capacitor (see Fig.~\ref{dovetail}a). The total inductance $L_{tot}$ is given by the sum of the two individual inductors $L_{tot} = L_1+L_2$ (see Fig.~\ref{dovetail}b). The dovetail inductor geometry is determined at the end of the design process, allowing the resonant frequency of the $LC$ circuit to be finalized once all capacitances have been determined. Each inductor possess a final geometry with a 120\um~stub width, which equals the width of both interdigital capacitors. The stubs fans out at standard 60$^{\circ}$ to a final radius of 1.8\mm~to achieve an inductance of 440\nH. 

\begin{figure}[t!]
  \begin{minipage}[c]{0.60\textwidth}
    \includegraphics[width=\textwidth]{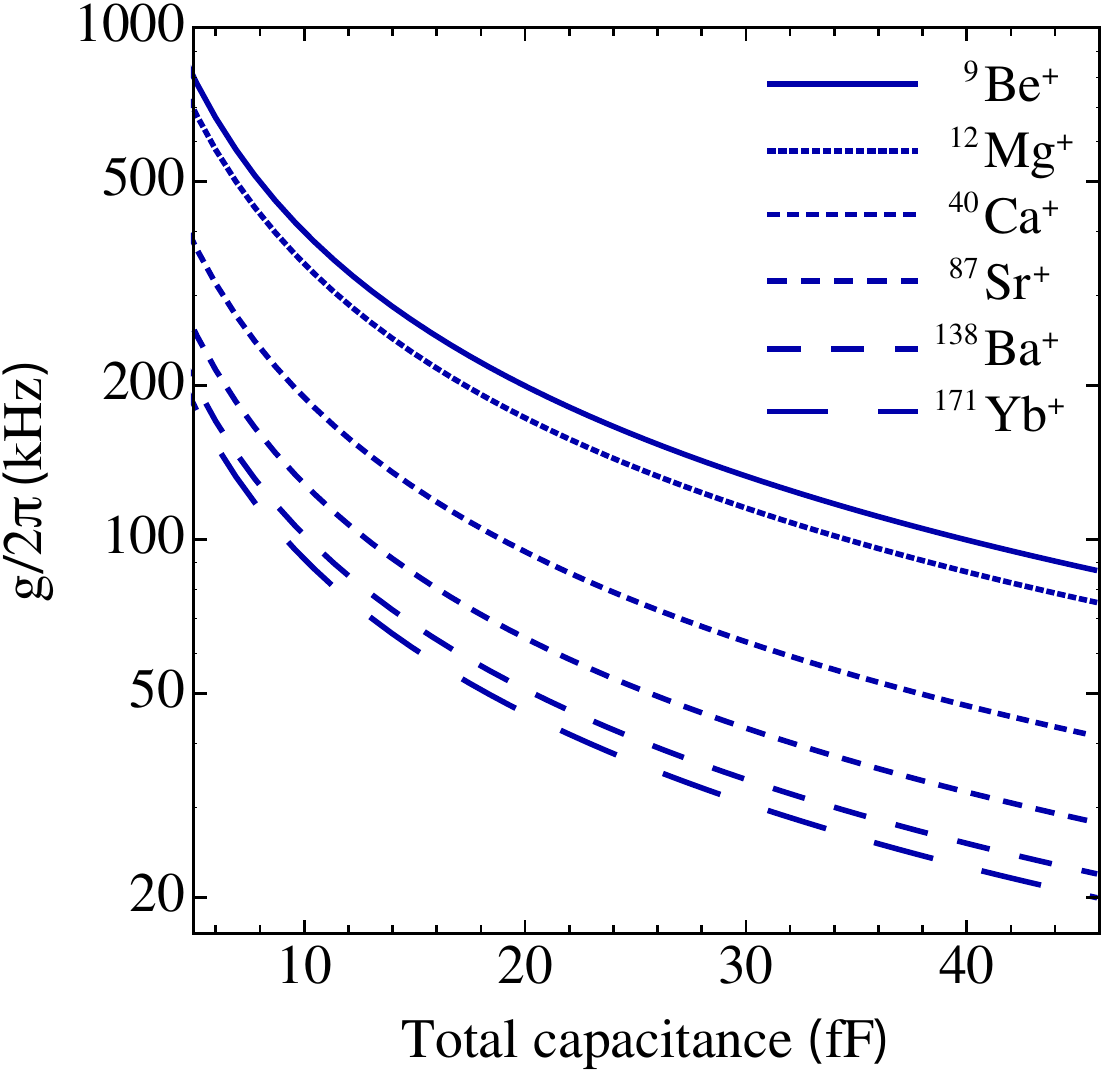}
  \end{minipage}\hfill
  \begin{minipage}[c]{0.35\textwidth}
    \caption{Coupling strength as a function of total circuit capacitance for $^{9}$Be$^+$, $^{12}$Mg$^+$, $^{40}$Ca$^+$, $^{87}$Sr$^+$, $^{138}$Ba$^+$ and $^{171}$Yb$^+$. The decrease in total capacitance by minimizing parasitic capacitances throughout the circuit allows for the increase in coupling strength between the ion and the resonator.}
     \label{elements}
  \end{minipage}
\end{figure}

As mentioned previously, reducing the total static capacitance of the circuit will also increase the coupling strength between the ion and the resonator (see Eq. (\ref{equg})). By using a radial stub design, the parasitic capacitances of the complete system are reduced to $<$5\ff. Fig.~\ref{elements} shows the coupling strength for $^{9}$Be$^+$, $^{12}$Mg$^+$, $^{40}$Ca$^+$, $^{87}$Sr$^+$, $^{138}$Ba$^+$ and $^{171}$Yb$^+$ as a function of the total capacitance of the circuit. It can be seen that even for heavy ions, such as $^{171}$Yb$^+$, a coupling strength of $g/2\pi \sim 200$\kHz~can be achieved by reducing the total circuit capacitance to $\sim$5\ff. It should be noted that as we decrease the total capacitance of the resonating system, we will also need to increase the length of the inductor to keep the same resonant frequency. Although increasing the length of the stub increases the capacitance to ground, the increased length causes a larger inductance change per unit length due to the wave propagation in the stub. The net result is that the capacitance formed by stub can be ignored as it is cancelled by the inductive component.


%
%
%
%
%
\subsection{Microfabrication}
\begin{figure}[t!]
\includegraphics[width=1\textwidth]{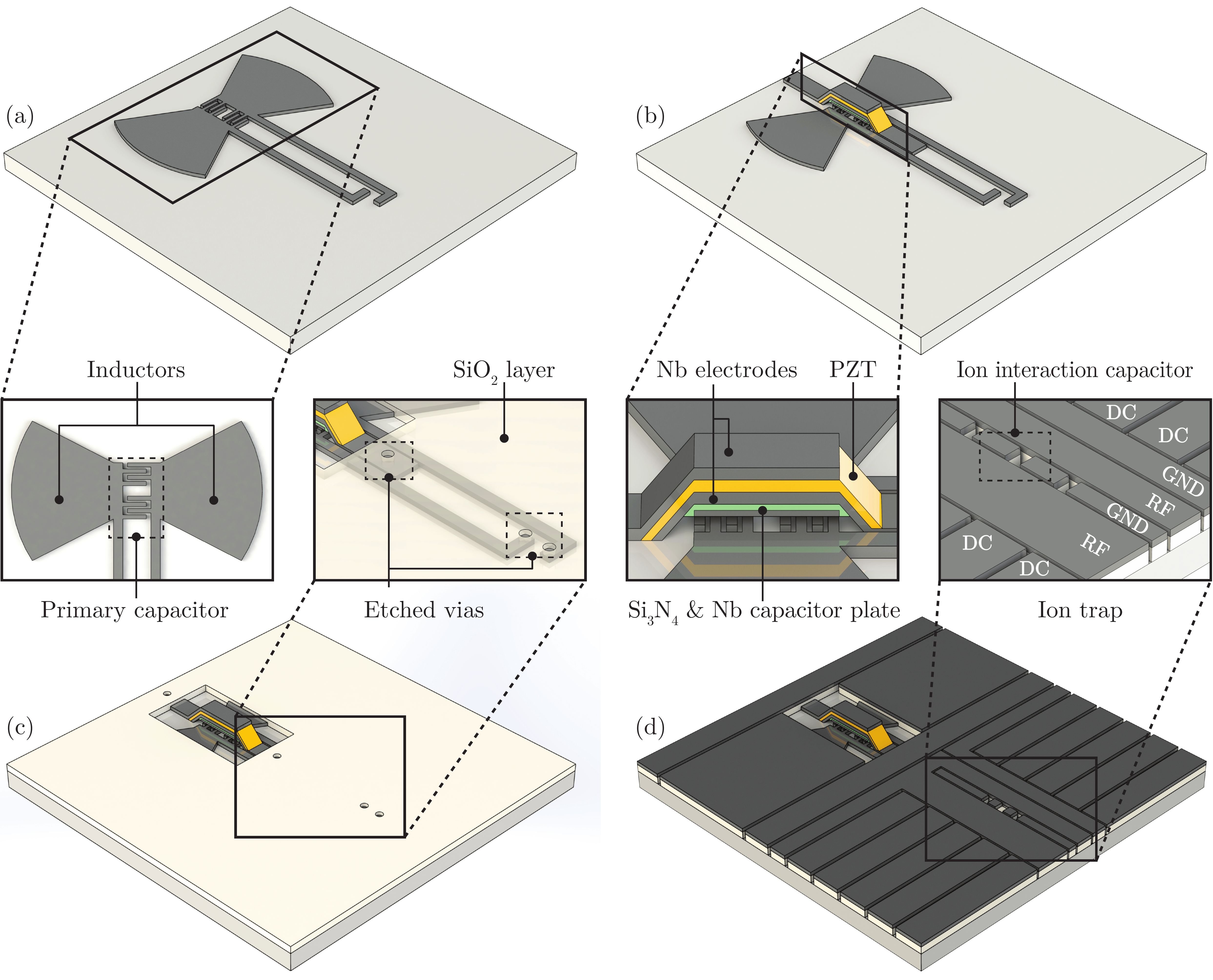}
\caption{Microfabrication process of the BAW device and ion trap structure. (a) Steps 1-2: Deposition and etching of the first Nb layer forming the $LC$ circuit. (b) Steps 3-10: Formation of the BAW device with the deposition and etching of a SiO$_2$ sacrificial layer, a Nb layer for the modulating capacitor plate, a Si$_3$N$_4$ insulation layer, two Nb BAW electrodes layers and the piezoelectric PZT layer. (c) Steps 11-12: Deposition of a SiO$_2$ insulating layer and etching to leave access for vias. (d) Steps 13: Final deposition of Nb with etching through the Nb and SiO$_2$ to produce the surface electrode structure.}
\label{proxy9}
\end{figure}

An overview of the microfabrication process for the ion trap electrodes and BAW device is presented in Fig.~\ref{proxy9}. The process was designed to minimize fabrication difficulties and processing time by limiting the number microfabrication techniques performed. To this end, all deposition techniques described are performed using a sputter deposition process and all dry etching processes are performed by inductively coupled plasma (ICP) etching. The following process steps have been numbered to help distinguish between the different stages of the microfabrication procedure.

\subsubsection{The BAW structure}

(1) The process begins with the preparation of a $250-500$\um~thick Al$_{2}$O$_{3}$ sapphire wafer, which is polished to a surface roughness of $<$0.5\nm~on both sides. Both surfaces are cleaned with acetone, methanol and isopropanol, followed by baking to dehydrate the surfaces. Further cleaning is performed using fuming nitric acid (FNA), a strong oxidizer that effectively removes organic material, ensuring a clean surface to fabricate upon. 

(2) The first layer deposited is a 1\um~Niobium (Nb) layer, which forms the interdigital capacitor and inductor of the superconducting $LC$ circuit. Prior to etching the Nb layer and all subsequent layers, a photolithography process is performed. A photosensitive material (photoresist) is spin coated on top, exposed to UV radiation through a mask and developed to form the desired pattern. The Nb layer can then be etched using a dry etching process.

(3) A 2\um~thick silicon dioxide (SiO$_{2}$) layer is then deposited. The layer will be used as a sacrificial layer that will be selectively removed later to form the air gap beneath the BAW device. (4) To ensure a flat surface for the deposition of subsequent layers which form the BAW device, the SiO$_{2}$ undergoes chemical mechanical polishing (CMP) to reduce surface roughness to under 10\nm. (5) A second 500\nm~thick Nb layer, which will serve as the moving plate that varies the capacitance of the circuit is then deposited, patterned and dry etched.

(6) A 500\nm~silicon nitride (Si$_{3}$N$_{4}$) layer is then deposited to provide an electrically insulating layer in between the capacitive plate and bottom Nb electrode used for the BAW. The Si$_{3}$N$_{4}$ and SiO$_{2}$ can then be dry etched to form the region underneath the BAW device.

(7) The first BAW electrode layer of 1000/100\nm~thick Niobium/Platinum (Pt) is deposited and dry etched. The Pt layer serves as an adhesion layer and diffusion barrier between the Nb and PZT layer. (8) The deposition of PZT can be performed through several methods, such as sol-gel coating \cite{Budd1985sol}, sputtering \cite{wilke2013sputter}, hydrothermal deposition \cite{morita1998cylindrical}, screen printing \cite{thiele2001processing} and aerosol deposition \cite{wang2006microfabrication}. For a 3\um~layer, rf magnetron sputtering provides an established deposition technique, which can utilize the same deposition tools as used for the other layers. (9) After patterning and dry etching the PZT layer, the second BAW electrode layer of 100/1000\nm~thick Pt/Nb is deposited and dry etched.

(10) A buffered hydrofluoric (BHF) acid wet etch is then used to etch the sacrificial SiO$_{2}$ layer. This step selectively removes the SiO$_{2}$ layer and provides the air gap below the BAW structure without attacking the surrounding Nb and Si$_{3}$N$_{4}$ layers.


\subsubsection{The ion trap electrodes}
(11) Once the BAW device has been fabricated, the ion trap electrode structures can then be deposited. The first layer deposited is a 3\um~thick SiO$_{2}$ layer used to insulate the ground electrodes of the $LC$ circuit from the top electrodes. The requirement of this insulating layer means fabricating the ion trap electrode structures must come after the BAW device, as any prior deposition of SiO$_{2}$ would be removed by the previous BHF wet etch. This process is made simpler as the ion trap electrodes and BAW device can be spatially separated on the chip. To prevent depositing subsequent layers over the BAW structure, thick photoresist layers ($\sim$30\um) can be used to protect covered areas. 

(12) The SiO$_{2}$ layer is then dry etched to provide space for vertical interconnects (vias) between the $LC$ circuit and the top electrode layer. (13) Finally, a 1\um~thick Nb layer is deposited to form the top electrode layer. The Nb and SiO$_{2}$ layer are then dry etched to form the electrode structure. 




\section{Ion trapping in a dilution refrigerator}\label{dilsec}

Ion trapping experiments possess several requirements that need to be adapted to a dilution fridge environment. Firstly, the utilization of a UHV system is necessary to minimize collisions between ions and residual background gases. For this purpose a dilution refrigerator and its cryogenic environment is ideal. When cooled down to mK temperatures it acts as a sorption pump and effectively pumps the cryostat to pressures unobtainable by room temperature vacuum systems. 

The preparation of single ions is traditionally performed by ohmically heating an atomic source material in an oven, followed by photoionization. However, within a dilution refrigerator this would be not possible due to insufficient cooling power at the sub-Kelvin stages. For low temperature ($\sim$4\K) ion trapping systems, alternatives such as laser ablation \cite{antohi2009} and photonic crystal fibres have already been demonstrated \cite{renn1995} and show potential for use in dilution fridges.

To generate electromagnetic trapping potentials, rf voltages of $\sim$100\V~and dc voltages of $<$10\V~are applied to trap electrodes.  Superconducting materials must also be used inside the setup to prevent the increase in temperature due to ohmic heating. A suitable material for trap structures is Nb (as discussed in section~\ref{bawsec}), which has a high critical temperature, critical magnetic field and critical current density. Electrical connections to the trap from outside of the cryostat can be achieved using commercially available NbTi coaxial cables. 

Cryostats for ion trapping applications use UHV compatible anti-reflection coated windows for optical access. These windows are coated to filter all wavelengths except those necessary for the experiment. However, dilution refrigerators use six radiation shields to prevent radiation from entering the cryostat to maintain sub-Kelvin temperatures. Therefore, a suitable alternative for the dilution refrigerator is the use of optical fibres and optical feedthroughs compatible with UHV and cryogenic environments. This solution also requires aligning devices such as low temperature piezo elements, which are capable of moving the fibre output within the dilution refrigerator.

Another challenge is the read out of information from the ion through the detection of fluorescence. Typically in ion trap experiments, the fluorescence is collected through an optical viewport using a lens system placed within a few centimetres of the ion trap. Photons collected are detected using either a charge coupled device (CCD) or a photomultiplier tube (PMT). As direct optical access is not desired, an alternative in-vacuum solution for cryogenic environments is a kinetic inductance detector (KID). Traditionally used for astronomical applications, KID technology was developed approximately 15 years ago, and shows results comparable to conventional CCD detectors \cite{mazin2013}.

Finally, care must also be taken when dealing with vibrations of the system. If liquid Helium (He) is used for pre-cooling the dilution refrigerator, vibrations will be negligible. However, operation of this kind of ``wet" refrigerator is difficult due to the requirement of adding liquid He continuously. Alternatively, a cryocooler can be used, which makes use of bellows to mechanically decouple the cryocooler from cryostat \cite{dean1999}.


\subsection{Ultra-high vacuum system} 
The first challenge in transferring a room temperature ion trapping infrastructure into a dilution fridge is optical access to the experiment. Although optical access to the 100\mK~stage in a dilution fridge has been demonstrated \cite{jessen2013}, optical fibres can be used to reduce heating into the system. As six radiation shields are used within the dilution refrigerator, the use of cryogenic compatible optical fibres and optical feedthroughs becomes inevitable to mitigate losses in laser power incurred at the windows on each shield. 

Cryogenic and UHV compatible optical feedthroughs are commercially available with specifications that are technically favourable for the proposed methodology. Such feedthroughs can be baked up to 200\degC~and cooled down to 3\K, with an attainable vacuum pressure of $1\times10^{-10}{\torr}$ at room temperature. A lower temperature limit of 3\K~is sufficient for these feedthroughs as they are attached to the vacuum shroud at room temperature. In addition, internal cabling connected to the feedthrough is first thermally anchored at the 40\K~cold stage. The specified transmission for optical fibre ranges between 370\nm~and 1200\nm, which is ideal for all wavelengths of the most commonly used atomic systems such as Ba$^{+}$, Yb$^{+}$, Sr$^{+}$ , Ca$^{+}$, and Mg$^{+}$, with a maximum attenuation of 1.2\dB/\m~for the shortest wavelength.


\subsection{Thermal anchoring}


\begin{figure}
  \begin{minipage}[c]{0.4\textwidth}
    \includegraphics[width=\textwidth]{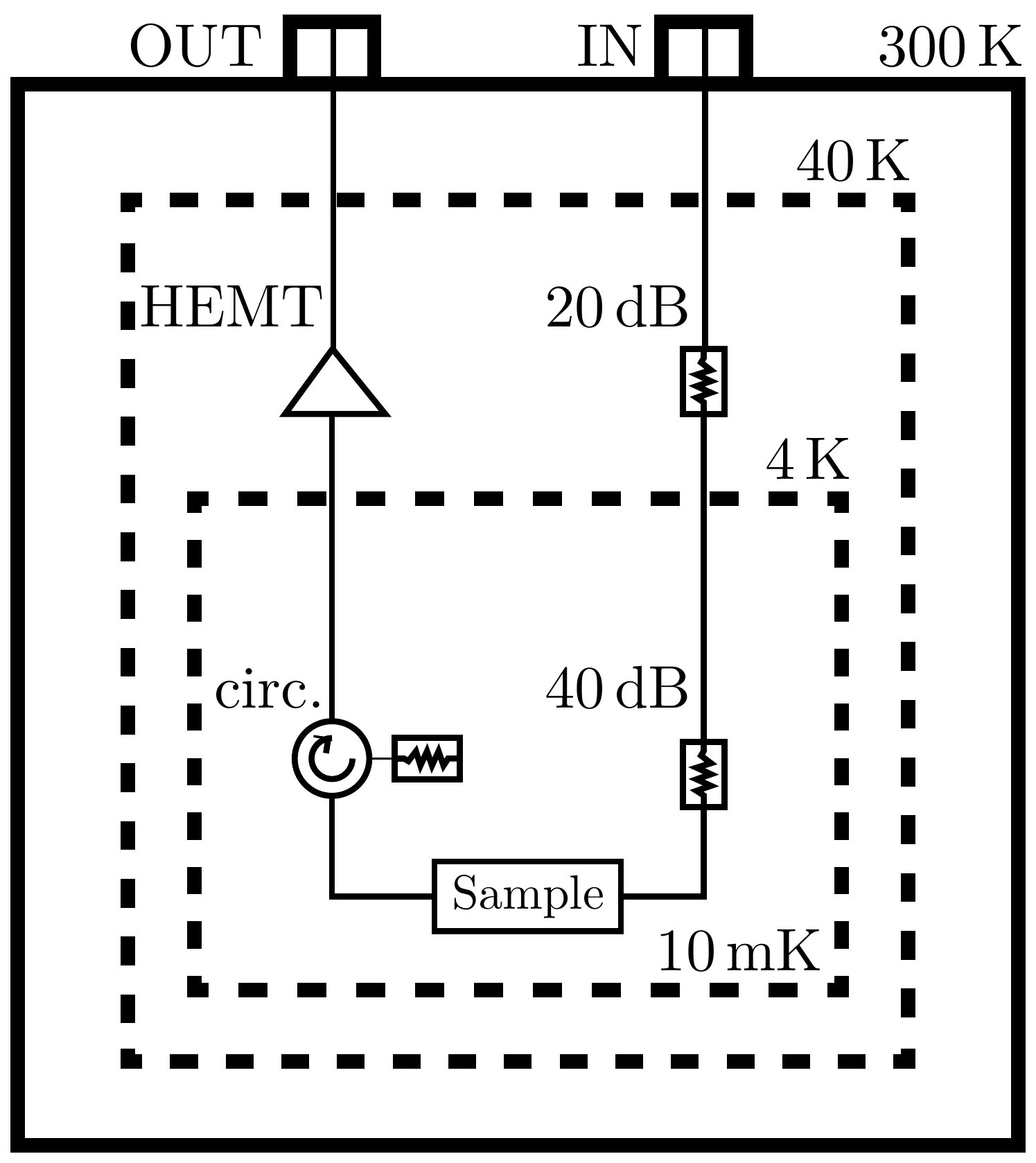}
  \end{minipage}\hfill
  \begin{minipage}[c]{0.5\textwidth}
    \caption{Microwave noise suppression setup for dilution refrigerator with base temperature of 10\mK. The input line shows two thermally anchored attenuators summing to 60\dB~to attenuate external noise. On the output line a thermally anchored cryocirculator and HEMT (high electron mobility transistor) cryoamplifier are also used to minimize noise.}
     \label{cryosetup}
  \end{minipage}
\end{figure}

\subsubsection{Microwave lines}

To avoid increasing the base temperature of the cryostat, the outer environment must be thermally decoupled from the internal environment. Coaxial lines for microwave and rf signals create thermal connections between the outer room temperature part of cryostat and the internal sub-Kelvin parts of the cryostat. To avoid unnecessary heating of the sample mounted on mixing chamber of the refrigerator, coaxial cables should be thermally anchored to multiple stages of the refrigerator. If the cables are thermally anchored, their temperature will be the same as the stage they are connected to. Therefore, we have a temperature gradient and heat flow through the cable. If the heat flow (the power coming into the cryostat) is higher than cooling power of refrigerator, the mixing chamber will not reach its base temperature. 

The power transmitted through the wire can be defined as $P = \lambda \delta T A / L$ \cite{ekin2006}, where A is the cross-sectional surface area, $L$ is the length, $\lambda$ is the mean thermal conductivity and $\delta T$ is the temperature difference across the length. A suitable choice for decreasing heat flow is to use conductors with lower $\lambda$. Commonly used materials for such applications are stainless steel or aluminium. For stages with lower temperatures, superconducting wires are also used (e.g. NbTi).

\subsubsection{Noise}

Thermal noise power can be determined from the Johnson-Nyquist formula $P_{th}=k_B T \nu_{BW}$ \cite{johnson1928, nyquist1928}, where $k_B$ is the Boltzmann constant, $T$ is the temperature in Kelvin and $\nu_{BW}$ is the frequency bandwidth in which the noise is measured. The noise power per hertz on an isolated chip on a 10\mK~mixing chamber is $1.38 \times 10^{-25}$\JHz. By connecting the chip to an external measurement system at $\sim$300\K~via coaxial cables, this number increases as the noise power of the outside system is $4.14\times 10^{-21}$\JHz, which is approximately $3.0\times 10^{4}$ times greater. To keep the noise at the mixing chamber low, external noise must be attenuated or filtered. For the input line, the ideal option is to use attenuators (in this case at least $60\dB \rightarrow 1.0\times 10^{6}$ times attenuation) that are thermally anchored to the stages of the refrigerator (see Fig.~\ref{cryosetup}). Attenuators attenuate both noise and input signal, however, the input signal can be increased by setting the output power on the microwave generator. For the output line, attenuators are not suitable; instead, we should use cryocirculators or HEMT (high electron mobility transistor) cryoamplifiers. As a passive device, cryocirculators can be mounted on any stage of the refrigerator, but preferably on the mixing chamber. Amplifiers on the other hand dissipate energy when operating and require cooling, and must therefore be placed on the stage with a high cooling power (4\K~stage cooled by pulse tube in dry refrigerators or by liquid He in wet refrigerators).

\subsubsection{DC lines}
DC wires must also be thermally anchored to the different stages of the refrigerator. To decrease the thermal coupling between the outside of the cryostat (300\K) and mixing chamber (10\mK), thermally resistive wires should be used (e.g. constantan, manganin, beryllium-copper or phosphor-bronze). Alternatively, if this is not possible, copper wires with a cross-section as small as possible should be used. For ion trap applications, the effect of ohmic heating, due to the wires, is not an issue as there is a very small current flowing through the DC electrodes and the resistances of the wires are only of the order of tens of ohms.


\subsection{Atomic sources}

Ohmically heated ovens, which are usually located within a few centimetres of an ion trap, are easy to construct, operate and offer a reliable method of delivering a specific atomic species to the trap. However, for cryogenic ion trapping applications these sources are unsuitable as they can dissipate between $10-20$\W~of power into the environment \cite{altafh.nizamani2011}. It has been shown that power dissipation at the coldest stage can be minimized by locating the atomic oven remotely in a warmer region of a cryostat \cite{sage2012}. The atoms are confined using magneto-optical traps, laser cooled and then pushed towards the ion trap using resonant laser light where they are finally photoionizied and trapped.

An alternate approach that has also been demonstrated is laser ablation, where an ion trap can be successfully loaded with a single 1.5\mJ~pulse \cite{Zimmermann2012}. This is a considerable benefit for systems operating at $\sim$4\K, however, it is still a significant heat load for dilution fridges that can only extract up to $\sim$10 mW~at 100\mK. 

Here we focus on the use of photonic crystal fibres which provide the substantial benefit of being a simple and demonstrated technology with the ability to load atoms externally and transport them to the experimental chamber with minimum heat load.

\subsubsection{Photonic crystal fibres (PCFs)}

\begin{figure}[t!]
  \begin{minipage}[c]{0.65\textwidth}
    \includegraphics[width=\textwidth]{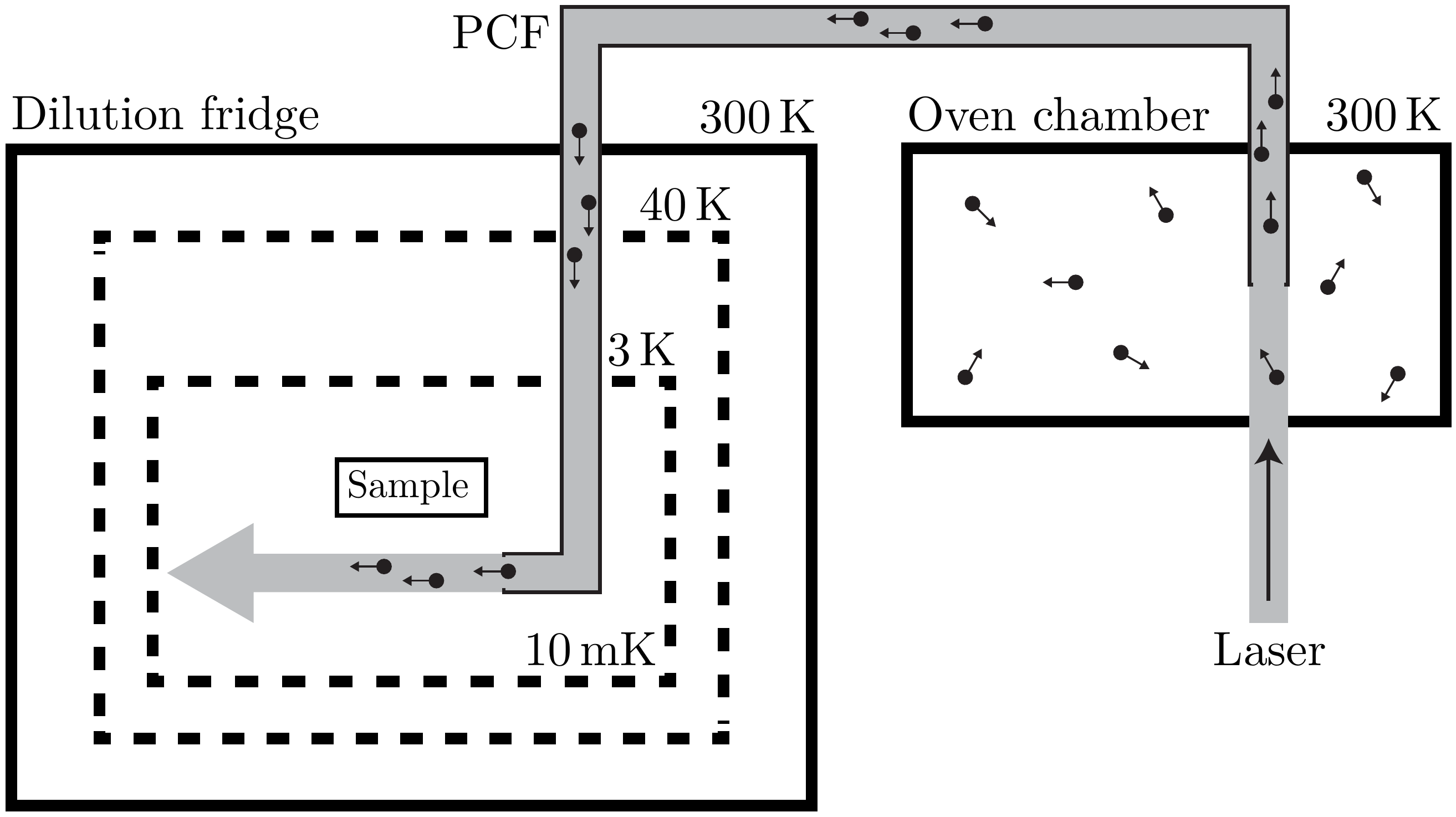}
  \end{minipage}\hfill
  \begin{minipage}[c]{0.3\textwidth}
    \caption{Resonant laser light is used to transport atoms from an external oven chamber, filled with an atomic cloud, into the dilution fridge using a photonic crystal fibre (PCF).}
     \label{pcfsetup02}
  \end{minipage}
\end{figure}

Optical fibres with hollow cores have been used to transport both light and atoms \cite{renn1995, ito1996}. For atom transport along hollow capillaries, the optical dipole forces of a the laser beam prevent adhesion to the surfaces and also provides the acceleration needed to overcome viscosity \cite{renn1995}. It has been shown that small dielectric particles can be trapped, controlled, and propelled in a laser beam using the forces exerted by light \cite{ol1993}. Fibres have also been shown to maintain UHV even with segments of the fibre exposed to atmospheric pressure and that they are capable of steering atoms through bends of the flexible fibre \cite{renn1995}. The fibres can therefore be used to transport atoms from an isolated system to the ion trap. This system can be located either at a warmer stage within the dilution fridge (4\K~or 40\K~stage) or outside of the fridge. 

As hollow core fibres are based on hollow capillaries to guide light in multiple modes, this can result in losses due to heterogeneous fields causing the destabilization of local fields for atomic transport. PCFs are a developing class of hollow optical fibres capable of transporting atoms using a single low loss spatial mode. PCFs are based on the properties of photonic crystals, which have a periodic optical nanostructure and rely on the photonic band gap principle to confine light allowing the fibre to have a hollow core \cite{cregan1999}. Transport of atoms has already been demonstrated using PCFs \cite{vorrath2010efficient}.

In table \ref{heattable} we summarise main sources of heat in the system, taking typical values for ion trapping laser powers and wire gauges. We have estimated the heat loading due to the DC connections using phosphor bronze wire, 36 AWG.

\begin{table}[h]
\centering
\begin{tabular}{|c||c|c|}
\hline
Source of heat & Estimated heat load & Heat sink location \\ \hhline{|=#=|=|}
Atomic oven & 20-30 W & 300 K Outside dilution fridge \\ \hline
DC wires & 3 mW & Multiple stages \\ \hline
Coax cables & 5 mW & Multiple stages \\ \hline
Laser beams & 1.8 mW & 4 K Stage \\ \hline
RF dielectric dissipation & 5 mW & 100 mK Stage \\ \hline
\end{tabular}
\caption{Estimated heat load and the location of the thermal anchoring.}
\label{heattable}
\end{table}



\subsection{Ion imaging} 

\subsubsection{Kinetic inductance detectors (KIDs)}

\begin{figure}[t!]
    \includegraphics[width=\textwidth]{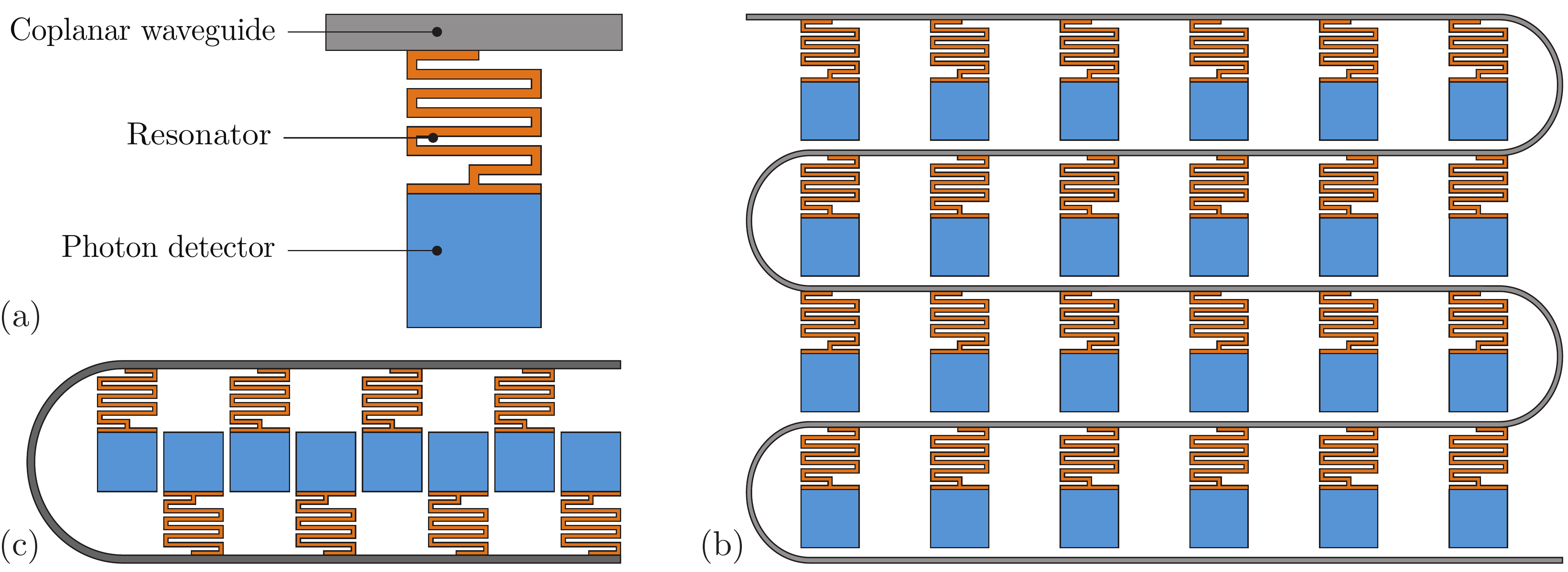}
    \caption{(a) Diagram of a $\lambda/4$ KID detector fabricated from a higher band gap superconductor. (b) Typical layout of KIDs in an array sharing a single CPW. (c) A linear high density KID array allowing for image resolutions of less than 10\um~in a single row.}
     \label{TopKID3}
\end{figure}

Traditional CCD or active pixel sensors, such as CMOS detectors, do not operate at cryogenic temperatures below 1\K. PMTs can also be used for ion detection \cite{gibbs1966large}, even in an array configuration. However, due to their size they are usually used for photon counting. One promising solution is a kinetic inductance detector (KID) (see Fig.~\ref{TopKID3}). A KID is a superconducting device that uses the effect of surface kinetic induction \cite{mazin,mazin2013,mattis1958} to change the resonant frequency, phase and amplitude of a high frequency microwave resonator. This corresponding frequency, phase and amplitude shift can be read out and interpreted as a photon flux density at the detector. 

There are several designs for KID detectors \cite{mazin,mazin2013,mchugh,roesch,roesch}, but the working principle is the same. A resonator structure is fabricated from a superconducting material, typically a $\lambda/4$ waveguide or lumped element design, which is tuned in the microwave frequency range ($10-20$\GHz). As photons of sufficient energy strike the superconductor they break up Cooper pairs, forming an excess of quasiparticles on the surface of the superconductor \cite{mazin,mazin2013}. These quasiparticles interact with phonons and cool rapidly forming a population with energies slightly greater than the gap energy and sub-gap phonons in the superconductor. These occupied energy states stop, by the Pauli exclusion principle, Cooper pairs from filling those same states, reducing the density of pairs \cite{mattis1958}. 

This has the effect of increasing the surface inductance of the superconductor and to a lesser extent the effective series resistance (ESR). As the superconductor is part of a resonant circuit, the increase in inductance and resistance lowers the frequency, phase and amplitude of the resonant peak. After a period of time the quasiparticles recombine into Cooper pairs when two interact and the resonator relaxes back into its original mode. 
The average distance the quasiparticles can travel before recombining is given by $l \approx \sqrt{D \tau_{qp}}$, where $\tau_{qp}$ is the quasiparticle lifetime and $D$ is the diffusion constant of the material, which is typically $>$8\pscm~for aluminium \cite{friedrich1997}. This recombination time is one of the components that determines the sensitivity of the detector.

To maximise the density of quasiparticles, reduce recombination time and separate the photon detector from the resonator structure, we can employ a quasiparticle trap which separates the quasiparticles and lowers their energy to below the gap energy of the detector superconductor. This can be accomplished by employing two dissimilar superconductors with different gap energies. If the photon strikes a Cooper pair in the higher gap superconductor, then the quasiparticles created will cool rapidly, via phonons, and flow into the lower gap superconductor. These quasiparticles will be unable to return to the high gap superconductor. 

\begin{figure}[t!]
\includegraphics[width=\textwidth]{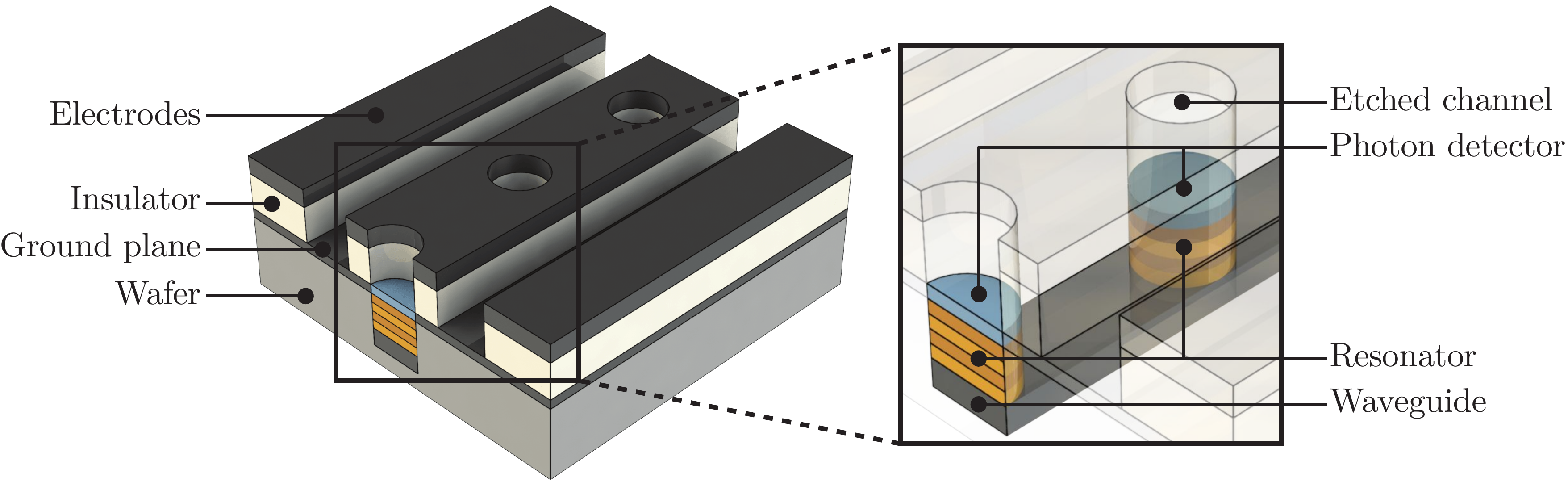}
\caption{Vertical KID structure featuring an into plane resonator. Light passes through etched channels in the electrode and strike the photon detector. The resonator is constructed from multi-plane elements that spiral down to the CPW, which is capacitively coupled to the resonator. The layer materials are Nb for the electrode surfaces, SiO$_{2}$ for the insulating layer and a Sapphire substrate.}
\label{ZoomedInTrap}
\end{figure}

Fig.~\ref{TopKID3} shows a typical $\lambda/4$ design where the photon detector is connected to the grounded end of the waveguide. The photon detector is the most sensitive point on a $\lambda/4$ design to quasiparticles and also forms the quasiparticle trap. To form an image, multiple KID cells must be multiplexed together in arrays. Since the resonant frequency of each cell can be tuned to a different frequency, we can address multiple KID detectors using a single wideband microwave waveguide. A 32 pixel \cite{roesch}, 144 pixel \cite{roesch}, 1024 pixel \cite{mchugh} and 2024 pixel \cite{mazin} KIDs have been fabricated and tested within dilution fridges and adiabatic demagnetization refrigerators at sub 1\K~temperatures. Progression has also been demonstrated to larger arrays with theoretical arrays of 100,000 pixels \cite{cancelo} being proposed. 

Fig.~\ref{TopKID3}b shows an example array created using multiple KID elements coupled onto a single coplanar waveguide (CPW) line. This topology is relatively simple to create but suffers from low pixel density as space is required for the CPW line and resonator structure between each photo detector. 

Fig.~\ref{ZoomedInTrap} shows a method of ion detection by placing the KID elements under the central electrode. This method allows for high density ion detection in a single line by etching holes separated by approximately $5\mu m$. The KID topology is formed between successive layers compared to Fig.~\ref{TopKID3} where the KID is created in a single layer.


\subsubsection{Flourescence detection}

For ion detection within a dilution fridge, we wish to limit the number of imaging feedthrough cables to reduce the thermal load on the system. The system would also need to have an inter pixel spacing of less than 10$\mu$m~so that neighbouring ions can be spatially resolved. These requirements make KID array detectors an ideal choice as designs featuring up to ten thousand pixels have been proposed \cite{vardulakis2008} using only two coaxial feedthrough lines. These arrays have pixel dimensions between $6-10 \mu$m.

While an external KID array and lens system could be designed, the KID network could also be incorporated within the trap structure itself. In such a design the Cooper pair photon detector can be constructed from the ground layer of the surface ion trap (see Fig.~\ref{ZoomedInTrap}). Each photon detector pad would be separated by a gap of approximately $2-3 \mu$m. Light from the ion would pass through etched channels in the trap structure down to the photon detector. 

Assuming the geometry of each detection pit to be $\approx7\um$ and $\approx10\um$ deep, which is achievable with microfabrication techniques, a numerical aperture of $\sim 0.75$ can be achieved depending upon the final geometry and allowing for adequate photon detection at the edge of each superconducting photon detector. We can estimate the cross-talk for the system by making the assumptions that we will only calculate edge coupled fields from adjoining photon detector sections including resonators and we will take the sum of the total cross-talk and ignore any directionality. If we have a frequency of 10 GHz on the photon detector and resonator with a separation of 5\um~along an edge length of 10\um~we get a cross-talk coefficient of -57.5 dB.

The quality factor of the resonator must be high such that each KID element can have a well-defined amplitude, frequency and phase component, but low enough that the signal strength is detectable. The signal from the waveguide would then be amplified by a HEMT or HFET (hetrostructure field-effect transistor) before being fed out for measurement. One design criteria of study is the usage of superconducting transistors for amplification purposes as this would allow amplification closer to the trap in the sub-Kelvin temperature range. This would allow for smaller noise floors and increases the potential pixel density for a single waveguide.

\begin{figure}[t!]
  \begin{minipage}[c]{0.4\textwidth}
    \includegraphics[width=\textwidth]{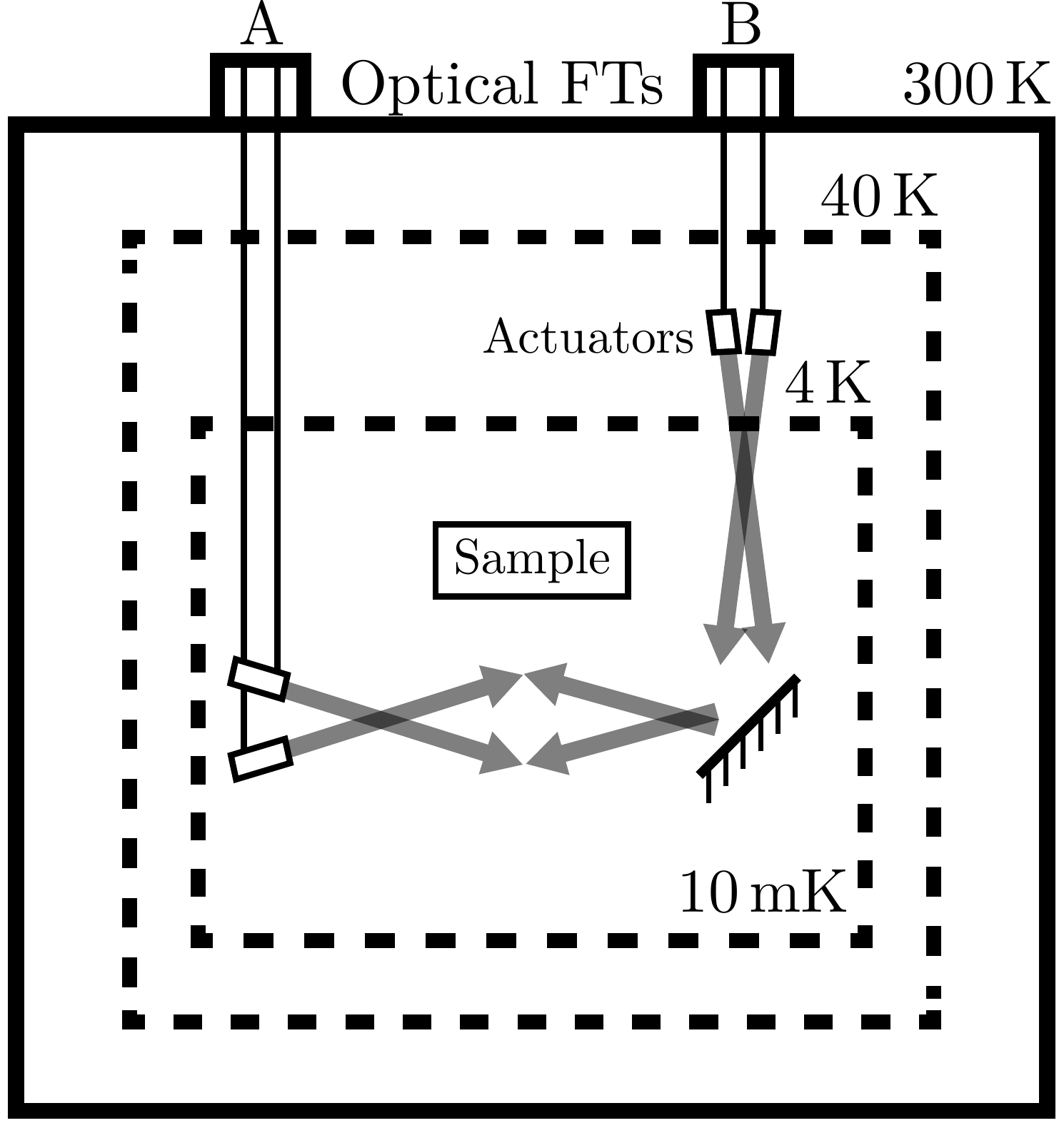}
  \end{minipage}\hfill
  \begin{minipage}[c]{0.5\textwidth}
    \caption{Two proposed methods for motion control of optical components within a dilution refrigerator. (A) Direct motion control at the 10\mK~stage. This method of control is suitable for both optical and photonic fibres with full independent translational and rotational control, however, it suffers from a reduced range of motion caused by low sub-Kelvin temperatures. (B) Direct motion control at the 4\K~stage.}
     \label{motion2}
  \end{minipage}
\end{figure}

\subsection{Motion control}

The need to position devices operating within the dilution fridge brings multiple challenges. The actuator must be capable of withstanding cryogenic temperatures below 1\K, stable under UHV and capable of rotational and translational motion at the micrometre scale. Commercially available ceramic actuators can fulfil these requirements and offer displacements between 6.5\um~and 32\um~with a blocking force of 190\N~to 3800\N. These devices have also been stress tested \cite{bosotti2005pi} and shown to work even after being placed under ten years of simulated thermal and structural stresses. However, it should be noted that these displacements are only quoted for liquid nitrogen temperatures at 77\K. Studies have also been performed on piezo actuators at sub-Kelvin temperatures, which show that while they continue to operate, they experience reduced expansion and flex motions \cite{fouaidy2005,fouaidy2007characterization}. 

\begin{figure}[t!]
    \includegraphics[width=\textwidth]{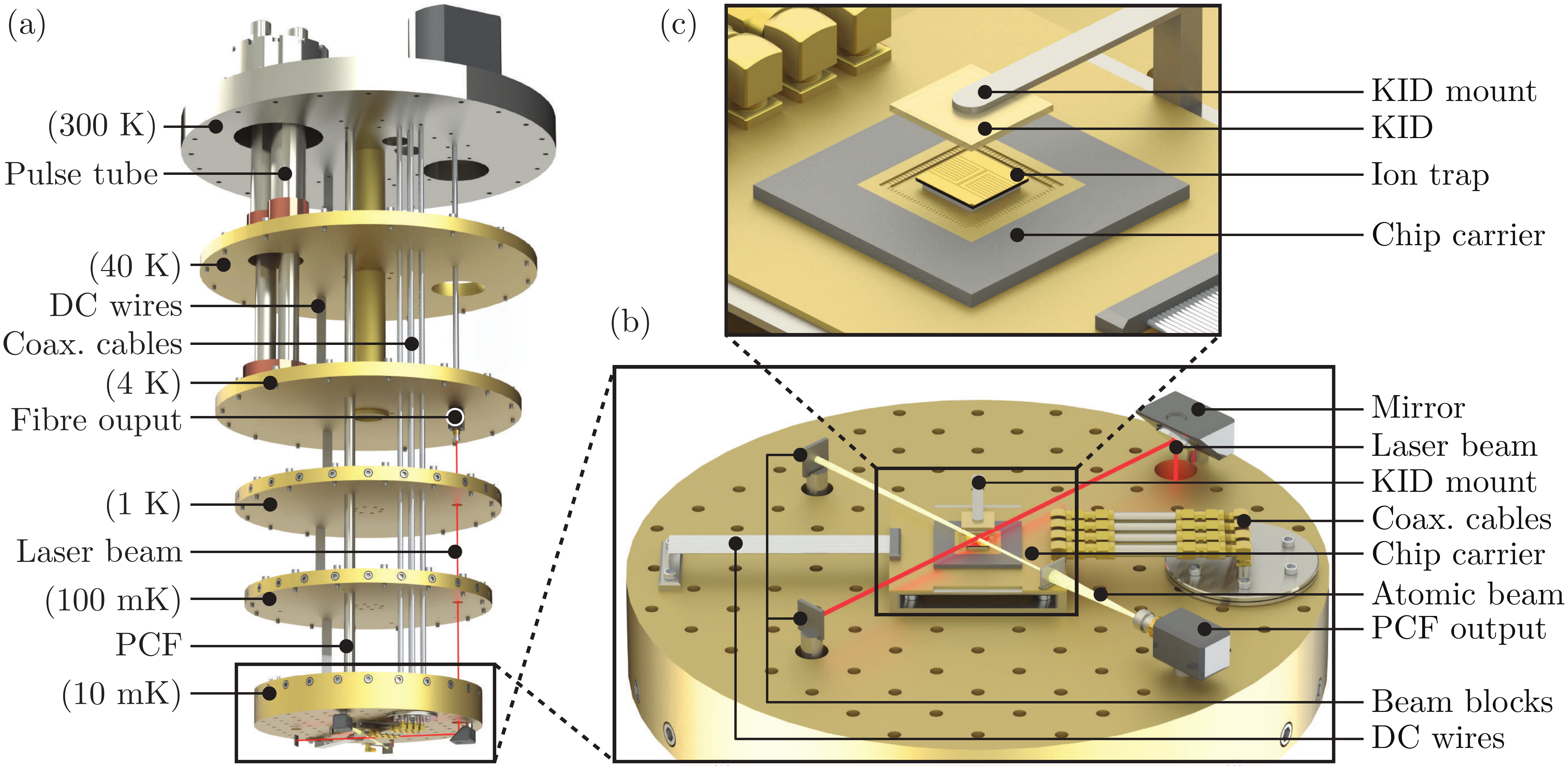}
    \caption{Final experimental layout of the ion trapping infrastructure within the dilution fridge environment. (a) The dilution fridge with the multiple temperature stages, cabling and laser access highlighted. (b) Mixing plate showing the ion trap and surrounding ancillary equipment. The Beam blocks are mounted on posts through the mixing plate and anchored to the 4K stage which is capable of 1.5W cooling power. (c) A magnified view of the ion trap position showing the KID mounted above the ion trap.}
     \label{proxy30}
\end{figure}

Utilizing piezoelectric actuators, there are two possible schemes for motion control of the fibre outputs. The first design, shown in feedthrough A of Fig.~\ref{motion2}, shows the motion controller positioned directly on the coldest stage of the dilution fridge. This provides direct translational and rotational control of the fibre. However, the dissipated heat and vibrational noise is positioned close to the ion trap and at millikelvin temperatures the actuators motion is limited to the micrometre regime \cite{fouaidy2005}.

The second design, shown in feedthrough B of Fig.~\ref{motion2}, is a hybrid between direct motion control and window access. This allows the positioning of the piezo actuators on a warmer stage, so that their thermal and vibrational noise is kept away from the coldest stage and their motion is maximized. However, in this design a mirror is needed to direct the beam towards the trap, thereby coupling the translational and rotational motion of the fibre together. This design would only work for the optical fibres as the atomic beam from the PCFs cannot be simply reflected.

Fig.~\ref{proxy30} shows an image of the dilution fridge with cabling and laser access highlighted. It also presents a proposed experimental layout on the mixing plate showing the ion trap, mounting scheme, electrical connections, laser access and delivery of the atomic source.



\section{Conclusion}

We have described the experimental methodology for the integration of an ion trapping and superconducting qubit system as a step towards the realization of a quantum hybrid system. Building upon work by Kielpinski et al. \cite{kielpinski2012}, we describe the design, simulation and microfabrication process of an ion trap capable of coupling the quantum state of an ion to a superconducting microwave $LC$ circuit with a coupling strength in the tens of kHz. We also discuss the challenges and difficulties in introducing an ion trapping experimental infrastructure into a dilution fridge, such as achieving UHV, maintaining mK temperatures, providing atomic sources, ion imaging, laser access and motion control, and present solutions that can be immediately implemented using current technology.

While many of the individual building blocks of the quantum hybrid system presented here have been demonstrated, it should be noted that the realization of such a system is an immense project to undertake, bringing together state-of-the-art technology platforms in multiple fields of science and technology. To reach the point of a fully operational system, where all components and systems are comprehensively optimized, will be a challenging task. However, here we provide an experimental system design that may indeed be experimentally realizable.


\begin{acknowledgements}

This work is supported by the U.K. Engineering and Physical Sciences Research Council [EP/G007276/1, EP/E011136/1, the UK Quantum Technology Hub for Networked Quantum Information Technologies (EP/M013243/1), and the UK Quantum Technology Hub for Sensors and Metrology (EP/M013243/1)], the European Commissions Seventh Framework Programme (FP7/2007-2013) under Grant Agreement No. 270843 (iQIT), the Army Research Laboratory under Cooperative Agreement No. W911NF-12-2-0072, U.S. Army Research Office Contract No. W911NF-14-2-0106, the Slovak Research and Development Agency Contracts No. DO7RP-0032-11, No. APVV-0515-10, and No. APVV-0808-12 and the University of Sussex. The views and conclusions contained in this document are those of the authors and should not be interpreted as representing the official policies, either expressed or implied, of the Army Research Laboratory or the U.S. Government. The U.S. Government is authorized to reproduce and distribute reprints for Government purposes notwithstanding any copyright notation herein.

\end{acknowledgements}

\bibliography{Hybridbib}

\end{document}